\documentclass[12pt,leqno]{amsart} %asmart
\usepackage{amssymb}
\usepackage{amsmath,epsfig,graphicx,color, float}
\usepackage{subcaption}
\usepackage{relsize}
\usepackage{comment}
\usepackage{hyperref}
\usepackage{xurl}
\usepackage{pdfpages}
\usepackage{graphicx}
\usepackage{mwe}
\usepackage{bbm}
\usepackage{algorithm}
\usepackage{algorithmicx}
\usepackage{algpseudocode}
\usepackage{listings}
\usepackage{booktabs}
\usepackage{fancyhdr}
\usepackage{longtable}
\usepackage{xcolor}
\usepackage{numprint}
\usepackage[export]{adjustbox}
\usepackage{wrapfig}

\usepackage[margin=1.05in]{geometry}
\hypersetup{
    colorlinks,
    linkcolor={blue!90!black},
    citecolor={blue!90!black},
    urlcolor={blue!90!black}
}

\numberwithin{table}{section}
\numberwithin{figure}{section}
\numberwithin{equation}{section}

\definecolor{darkblue}{rgb}{.2, 0.2,.8}
\definecolor{darkgreen}{rgb}{0,0.5,0.3}
\definecolor{darkred}{rgb}{.8, .1,.1}

\newcommand{\bfY}{\vect{Y}}

\newcommand{\bfalp}{\vect{\alpha}}

\newcommand{\bfT}{\mat{T}}
\newcommand{\bft}{\vect{t}}
\newcommand{\bfe}{\vect{e}}
\newcommand{\bfpi}{\vect{\pi}}
\newcommand{\bfBeta}{\boldsymbol{\beta}}
\newcommand{\bfp}{\vect{\pi}}

\newcommand{\calL}{\mathcal{L}}
\newcommand{\calT}{\mathcal{T}}

\newcommand{\0}{\mat{0}}

\newcommand{\E}{\mathbb{E}}
\renewcommand{\P }{{\mathbb P}}
\newcommand{\ci}{\mathrel{\text{\scalebox{1.07}{$\perp\mkern-10mu\perp$}}}}

\newtheorem{remark}{Remark}[section]

\usepackage{bm}
\newcommand{\vect}[1]{\pmb{#1}}
\newcommand{\mat}[1]{\boldsymbol{\bm #1}}

\DeclareMathOperator*{\argmax}{arg\,max}

\newcommand{\Ti}{\mat{T}_i}
\newcommand{\Tl}{\mat{T}_l}

\newcommand{\ti}{\vect{t}_i}
\newcommand{\tl}{\vect{t}_l}

\newcommand{\ek}{\vect{e}_k}
\newcommand{\ekt}{\vect{e}^{\top}_k}

\newcommand{\evec}{\vect{e}}

\newcommand{\ejt}{\vect{e}^{\top}_j}

\newcommand{\est}{\vect{e}^{\top}_s}

\newcommand{\xim}{x_{i}^{(m)}}
\newcommand{\xlm}{x_{l}^{(m)}}
\newcommand{\rc}{\delta_{i}^{(m)}}
\newcommand{\rcl}{\delta_{l}^{(m)}}
\newcommand{\probObs}{\sum_{j=1}^{p}\pi_{j}^{(m)} \prod_{i=1}^{d}\left(  \ejt  \exp \left(\Ti\xim \right)\ti \right)^{\rc}
    \left(  \ejt  \exp \left(\Ti\xim \right)\evec\right)^{1-\rc}}

\newcommand{\akMinusI}{\prod_{l\neq i}\left(  \ejt  \exp \left(\Tl\xlm \right)\tl \right)^{\rcl}
    \left(  \ejt  \exp \left(\Tl\xlm \right)\evec\right)^{1-\rcl}}

\newcommand{\matInt}{\int_{0}^{\xim}  \exp\left(\Ti (\xim-t)\right)\ti \ejt \exp(\Ti t)\,dt \,}

\newcommand{\matIntRc}{\int_{0}^{\xim}  \exp\left(\Ti (\xim-t)\right)\evec\ejt \exp(\Ti t)\,dt \,}

\allowdisplaybreaks

\begin{document}
\bibliographystyle{plain}

\title{Joint lifetime modelling with matrix distributions}

\author[H. Albrecher]{Hansj\"org Albrecher}
\address{Department of Actuarial Science, Faculty of Business and Economics, University of Lausanne, UNIL-Dorigny, 1015 Lausanne and Swiss Finance Institute, 1015 Lausanne}
\email{hansjoerg.albrecher@unil.ch}

\author[M. Bladt]{Martin Bladt}
\address{Department of Actuarial Science, Faculty of Business and Economics, University of Lausanne, UNIL-Dorigny, 1015 Lausanne}
\email{martin.bladt@unil.ch}

\author[A. M\"uller]{Alaric J.A. M\"uller}
\address{Department of Actuarial Science, Faculty of Business and Economics, University of Lausanne, UNIL-Dorigny, 1015 Lausanne}
\email{alaric.mueller@unil.ch}

\begin{abstract}
Acyclic phase-type (PH) distributions have been a popular tool in survival analysis, thanks to their natural interpretation in terms of ageing towards its inevitable absorption. In this paper, we consider an extension to the bivariate setting for the modelling of joint lifetimes. In contrast to previous models in the literature that were based on a separate estimation of the marginal behavior and the dependence structure through a copula, we propose a new time-inhomogeneous version of a multivariate PH class (mIPH) that leads to a model for joint lifetimes without that separation. We study properties of mIPH class members and provide an adapted estimation procedure that allows for right-censoring and covariate information. We show that initial distribution vectors in our construction can be tailored to reflect the dependence of the random variables, and use multinomial regression to determine the influence of covariates on starting probabilities. Moreover, we highlight the flexibility and parsimony, in terms of needed phases, introduced by the time-inhomogeneity. Numerical  illustrations are given for the data set of joint lifetimes of Frees et al.\ \cite{frees(1996)}, where 10 phases turn out to be sufficient for a reasonable fitting performance. As a by-product, the proposed approach enables a natural causal interpretation of the association in the ageing mechanism of joint lifetimes that goes beyond a statistical fit. 
\end{abstract}
\maketitle
%\makeatletter
%\enddoc@text
%\let\enddoc@text\empty % to remove the contact info from the end of the document
%\makeatother
%%%%%%%%%%%%%%%%%%%%%%%%%%%%%%%%%%%%%%%%%%%%%%%%%%%%%%%%%%%%%%%%%%
%----------------------------------------------------------------% 
%%%%%%%%%%%%%%%%%%%%%%%%%%%%%%%%%%%%%%%%%%%%%%%%%%%%%%%%%%%%%%%%%%
\section{Introduction}
When studying insurance products on multiple lives, it is natural to assume that individuals who are exposed to very similar life conditions may have somewhat correlated lifetimes. This is especially true for married couples, since, once married, the spouses typically share to a large extent a similar lifestyle. Indeed, the simplistic assumption of independence of lifetimes of partners has been shown to be inappropriate in various papers. For example, Frees et al.\ \cite{frees(1996)} used a bivariate Frank copula model to assess the effect of dependency between husband and wife on insurance annuities, illustrating their approach on a by now classical data set of a large insurer. 
The same data was used in Carriere \cite{carriere2000bivariate}, where multiple bivariate copula models were studied. The Linear-Mixing Frailty copula was found to be the best suited model to describe the data.
Shemyakin \& Youn \cite{shemyakin2006} introduce a general conditional Bayesian copula for joint last survivor analysis. They allow entry ages of spouses to have a selection effect on his/her mortality, as well as on the other spouse’s mortality. 
For that same data set, Luciano et al.\ \cite{luciano2008} captured the dependence between survival times of spouses by an Archimedean copula, whose marginals were estimated according to a stochastic intensity approach.  In a similar spirit, Dufresne et al.\ \cite{dufresne(2018)agediff} allow for the Archimedean copula parameter to depend on the age difference of partners at issue of the policy, in order to describe the dependence of the remaining lifetime of a couple. In Gobbi et al.\ \cite{gobbi_kolev_mulinacci_2019}, extended Marshall-Olkin models were employed for that same data set, where the continuous copula approach is extended by allowing for fatal events that affect both marginal lives.\\

In this paper we propose an alternative to copula-based methods for the modelling of joint remaining lifetimes in a couple based on multivariate phase-type distributions. Phase-type (PH) distributions are interesting candidates since they broaden favourable properties of exponential random variables to scenarios where the latter alone would not be appropriate. In particular, the denseness of PH distributions among all distributions on the positive half-line in the sense of weak convergence, which extends to the multivariate setup, is a major advantage when one wants to approximate a distribution. For more details on PH distributions we refer readers to Bladt \& Nielsen \cite{Bladt2017}. As opposed to copula-based methods, a PH distribution can give rise to a natural interpretation when used to approximate lifetime distributions. One can view the path of a Markov jump process as the life of an individual, which goes through several different states (for instance biological markers) before reaching the inevitable absorption (death) state. 
Along this interpretation, acyclic PH distributions have been the first choice for modelling the ageing process of a human life, since they only allow forward transitions or direct exits to the absorption state. This characteristic makes them an appropriate tool for describing lifetimes ended by natural ageing or accidents. In Lin \& Liu \cite{markovageing2007Lin&Liu}, a PH distribution with Coxian structure was used to explain the physical ageing process of marginal lifetimes. This approach was extended in Asmussen et al.\ \cite{Asmussen2019} to generalised Coxian distributions for the purpose of pricing equity-linked products. 
The first contribution to lift the PH approach to bivariate lifetime models was Ji et al.\ \cite{markovapproach2011}, where a Markovian multi-state model and a semi-Markov model are used to describe the dependence between the lifetimes of husbands and wives. Spreeuw \& Owadally \cite{spreeuw_owadally_2013} also use a Markovian multi-state model, with more attention given on how to tie the bereavement effect to forces of mortality. Moutanabbir \& Abdelrahman \cite{sarmanov2021} then used a bivariate Sarmanov distribution with PH marginals to model joint lifetimes. Both papers focused on the pricing of multiple-life insurance contracts.\\
Recently, Albrecher et al.\ \cite{ABBY21} introduced time-inhomogeneous PH (IPH) distributions for the purpose of lifetime modelling, which leads to a considerable reduction of necessary phases for a satisfactory fit of given data, since the introduced inhomogeneity can more efficiently accommodate non-exponential shapes than an augmentation of the phase dimension. In particular, \cite{ABBY21} applied regression on the intensity functions of the IPH distributions to associate lifetimes of different cohorts and populations. \\
In this paper, we propose a different route for using available information in the data set, namely to 
incorporate multinomial logistic regressions in the estimation procedure of multivariate PH distributions. In particular, the regression is applied to the initial distribution vectors of each IPH component, which adapts an approach presented in Bladt \& Yslas \cite{BladtYslas2022} to the multivariate case. The resulting dependence structure allows for explicit formulas alongside an intuitive `ageing' interpretation and, beyond the theoretical contribution, for a satisfactory fit to the bivariate spouses' lifetime data. \\

The remainder of the paper is structured as follows. Section 2 introduces the class of multivariate PH distributions that we will use to describe joint lifetimes of couples; we also provide some additional properties. In Section 3 an estimation method for this multivariate PH distribution is introduced, which allows for right censoring and covariate information. Section 4 then applies and illustrates the procedure on the classical spouses' lifetime data set from \cite{frees(1996)} and interprets the results. Section 5 concludes.  
%%%%%%%%%%%%%%%%%%%%%%%%%%%%%%%%%%%%%%%%%%%%%%%%%%%%%%%%%%%%%%%%%%
%----------------------------------------------------------------%
%%%%%%%%%%%%%%%%%%%%%%%%%%%%%%%%%%%%%%%%%%%%%%%%%%%%%%%%%%%%%%%%%%
\section{Multivariate phase-type distributions}
We first recall the mPH class, which was introduced in Bladt \cite{bladt2022tractable}.
\subsection{mPH distributions}
Let $ \{ J_t^{(i)} \}_{t \geq 0}$, $i=1,\dots,d$, denote separate homogeneous Markov pure-jump processes on the common state space $E=\{1, \dots, p, p+1\}$, with states $1,\dots,p$ being transient and $p+1$ absorbing. Defining transition probabilities as
\begin{align*}
p^{(i)}_{jl}(s,t)=\P(J_t^{(i)}=l|J_s^{(i)}=j)\,,\quad 0\le j,l\le p+1,\:\:0< i\le d \,,
\end{align*}
we may write
$$\mat{P}_i(s,t)=\exp(\mat{\Lambda}_i(t-s))=
\left( \begin{array}{cc}
		\exp(\bfT_i(t-s)) &  \bfe-\exp(\bfT_i(t-s)) \bfe \\
		\0 & 1
	\end{array} \right)\in\mathbb{R}^{(p+1)\times(p+1)},$$
for $s<t,$ $0\le i\le d$, where $\mat{\Lambda}_i(t)$ are intensity matrices.
In the following we write $\bfe_k$ for the $k$-th canonical basis vector in $\mathbb{R}^p$, $\bfe=\sum_{j=1}^p \bfe_j$, and
$$\bfT_i=\{t_{ks}^{(i)}\}_{k,s=1,\dots,p}\,,\quad \bft_i=-\bfT_i\bfe=(t^{(i)}_1,\dots,t^{(i)}_p)^{\mathsf{T}}\,,\:\:\: k=1,\dots,p\,.$$ 
%,\quad \lambda^{(i)}_k=-t_{kk}^{(i)}>0\,,
The crucial property of this class of PH distributions is now its dependence structure. Concretely, the assumption is that all jump processes start in the same state at time $t=0$, but proceed independently thereafter until absorption. That is, dependence is introduced solely through the shared initial state, which leads to a particularly tractable yet flexible model class. More formally, 
\begin{align}\label{dependence_def}
J_0^{(i)}=J_0^{(l)},\quad \{ J_t^{(i)} \}_{t \geq 0} {\ci}_{J_0^{(1)}} \{ J_t^{(l)} \}_{t \geq 0,\:l \neq i},\quad \forall i,l\in\{1,\dots,d\}.   
\end{align}
We will use $J_0:=J_0^{(i)}$ to simplify notation. Let $\P(J_0=j)=\pi_j$, $j=1,\dots,p$ and $\bfp=(\pi_1,\dots,\pi_p)$ denote the distribution vector of the shared initial state. The random variables
\begin{align}\label{components_def}
X_i = \inf \{ t >  0 : J^{(i)}_t = p+1 \}\,, \quad i=1,\dots,d,
\end{align}
are then all univariate PH distributed.
% and will be used later on to model remaining lifetimes of individuals $i$. 
% The distribution of  encloses the joint behaviour which will be our focus in the following, in particular its inhomogeneous extension. 
 We say that the random vector $X=\begin{pmatrix} X_1,\ldots, X_d \end{pmatrix}\in\mathbb{R}_+^{d}$ has a multivariate phase-type distribution (mPH) if each marginal variable $X_i$, $i=1,2,\dots,d$ is given by \eqref{components_def} and pairwise dependence is defined by \eqref{dependence_def}. We use the notation $$X\sim \mbox{mPH}(\bfp,\mathcal{T}), \quad \mbox{with}\quad \mathcal{T}=\{\bfT_1,\dots,\bfT_d\}.$$ 
% for the distribution of the random vector with characteristics defined above. 
The joint cumulative distribution function of $X$ is given by
\begin{align*}
 F_X(x)&=\P(X_1\le x_1,X_2\le x_2,\dots,X_d\le x_d)\\
       &=\sum_{j=1}^{p} \P(X_1\le x_1,X_2\le x_2,\dots,X_d\le x_d \mid J_0=j)\P(J_0=j)\\
       &=\sum_{j=1}^p\pi_j \prod_{i=1}^d\left(1-\bfe_j^\mathsf{T}\exp(\bfT_i x_i)\bfe\right),   \quad x\in\mathbb{R}_+^d. 
\end{align*}
%where we easily observe the independence of the components $X_i$'s once a condition on $J_0$ is introduced. By the same logic we have 
Furthermore, the survival function is
\begin{align*}
S_X(x)&=\P(X_1>x_1,X_2> x_2,\dots,X_d> x_d)\\
&=\sum_{j=1}^p\pi_j \prod_{i=1}^d\bfe_j^\mathsf{T}\exp(\bfT_i x_i)\bfe,
\end{align*}
and the probability density function is given by
\begin{align*}
f_X(x)&=\sum_{j=1}^p\pi_j \prod_{i=1}^d\bfe_j^\mathsf{T}\exp(\bfT_i x_i)\bft_i.
\end{align*}
For more details, cf.\ \cite{bladt2022tractable}.
%%%%%%%%%%%%%%%%%%%%%%%%%%%%%%%%%%%%%%%%%%%%%%%%%%%%%%%%%%%%%%%%%%
%----------------------------------------------------------------%
%%%%%%%%%%%%%%%%%%%%%%%%%%%%%%%%%%%%%%%%%%%%%%%%%%%%%%%%%%%%%%%%%%
\subsection{mIPH distributions}
The particular focus in this paper will now be on an inhomogeneous extension of the mPH distribution (briefly mentioned in \cite[Sec.6.1]{bladt2022tractable}). When considering time-inhomogeneous Markov pure jump processes on the common state-space $E$, it follows from Albrecher \& Bladt \cite{albrecher2019inhomogeneous} that the transition matrices are modified to
$$\mat{P}(s,t)=\prod_{s}^{t}(\boldsymbol{I}+\boldsymbol{\Lambda}(u) d u):=\boldsymbol{I}+\sum_{k=1}^{\infty} \int_{s}^{t} \int_{s}^{u_{k}} \cdots \int_{s}^{u_{2}} \mathbf{\Lambda}\left(u_{1}\right) \cdots \mathbf{\Lambda}\left(u_{k}\right) d u_{1} \cdots \mathrm{d} u_{k},$$
with sub-intensity matrix
\begin{align*}
	\mat{\Lambda}(t)= \left( \begin{array}{cc}
		\bfT(t) &  \bft(t) \\
		\0 & 0
	\end{array} \right)\in\mathbb{R}^{(p+1)\times(p+1)}\,, \quad t\geq0\,.
\end{align*}
The random variables $Y_i = \inf \{ t >  0 : J^{(i)}_t = p+1 \}\,, \: i=1,\dots,d,$ then follow univariate inhomogeneous phase-type (IPH) distributions, cf.\  \cite{albrecher2019inhomogeneous} for more details.\\
Here we focus on the particularly tractable case  $\bfT_i(t)=\lambda_i(t)\bfT$.  A random vector $Y=\begin{pmatrix} X_1,\ldots, X_d \end{pmatrix}$ is said to have an inhomogeneous multivariate PH (mIPH) distribution if all marginals follow IPH distributions and the dependence structure is defined by \eqref{dependence_def}. We write 
$$Y\sim \mbox{mIPH}(\bfp,\mathcal{T},\mathcal{L}), \quad \mbox{where}\quad \mathcal{T}=\{\bfT_1,\dots,\bfT_d\},\quad \mathcal{L}=\{\lambda_1,\dots,\lambda_d\}.$$
With
$$g^{-1}_i(y):=\int_0^y \lambda_i(u)du,\quad i=1,\dots, d,$$
the cumulative distribution function, survival function and density of $X$ are given by
\begin{align*}
F_Y(y)&=\sum_{j=1}^p\pi_j \prod_{i=1}^d(1-\bfe_j^\mathsf{T}\exp(\bfT_i g^{-1}_i(y_i))\bfe), \quad y\in\mathbb{R}_+^d,\\
S_Y(y)&=\sum_{j=1}^p\pi_j \prod_{i=1}^d\bfe_j^\mathsf{T}\exp(\bfT_i g^{-1}_i(y_i))\bfe,\quad y\in\mathbb{R}_+^d,\\
\end{align*}
and 
\begin{align*}
f_Y(y)&=\sum_{j=1}^p\pi_j \prod_{i=1}^d\bfe_j^\mathsf{T}\exp(\bfT_i g_i^{-1}(y_i))\bft_i\lambda_i(y_i),\quad y\in\mathbb{R}_+^d,
\end{align*}
respectively. Note that one can view each IPH random variable as a transformation of a PH random variable (and correspondingly the absorption time of a time-transformed formerly time-homogenous Markov jump process), with $X\sim \mbox{PH}(\bfpi,\bfT)$ and $g(X)\sim\mbox{IPH}(\bfpi,\bfT,\lambda)$.\\
The construction of  $\mbox{mIPH}(\bfp,\mathcal{T},\mathcal{L})$ allows different sub-intensity matrices and inhomogeneity functions for each marginal, as long as they share the same state-space. This leads to a considerable model flexibility. In particular, when compared to the homogeneous case, time-inhomogeneity allows for substantially smaller state-spaces for appropriate fits of data with potentially non-exponential tails (cf.\ \cite{albrecher2019inhomogeneous}), and the  mIPH class inherits this feature. 

When we condition a mIPH distribution on one or more marginals, we obtain another mIPH distribution with a new initial distribution vector and smaller dimension: 
Let $Y\sim \mbox{mIPH}(\bfp,\mathcal{T},\mathcal{L})$ and condition on the value of $Y_l$, $l\leq d$. The conditional density is 
$$
f_{Y \mid Y_l}(y\mid y_l)=\sum_{j=1}^p\frac{\pi_j\bfe_j^\mathsf{T}\exp(\bfT_l g_l^{-1}(y_l))\bft_l\lambda_l(y_l)}{\bfpi\exp(\bfT_l g_l^{-1}(y_l))\bft_l\lambda_l(y_l)} \prod_{i\neq l}\bfe_j^\mathsf{T}\exp(\bfT_i g_l^{-1}(y_i))\bft_i\lambda_i(y_i).
$$ 
That is,
\begin{equation}\label{cond_dist}
Y\mid Y_l=y_l \sim \mbox{mIPH}(\bfalp,\mathcal{T} \setminus \bfT_l,\mathcal{L} \setminus \lambda_l),
\end{equation}
with initial distribution vector
$$\bfalp=\left\{\pi_j\times \frac{\bfe_j^\mathsf{T}\exp(\bfT_l g_l^{-1}(y_l))\bft_l\lambda_l(y_l)}{\bfpi\exp(\bfT_l g_l^{-1}(y_l))\bft_l\lambda_l(y_l)}\right\}_{j=1,\dots,p}.$$ The same reasoning can be applied to obtain 
\begin{equation}\label{cond_dist_surv}
Y\mid Y_l\geq y_l \sim \mbox{mIPH}(\vect{\nu},\mathcal{T} \setminus \bfT_l,\mathcal{L} \setminus \lambda_l),
\end{equation} with 
$$\vect{\nu}=\left\{\pi_j\times \frac{\bfe_j^\mathsf{T}\exp(\bfT_l g_l^{-1}(y_l))\bfe}{\bfpi\exp(\bfT_l g_l^{-1}(y_l))\bfe}\right\}_{j=1,\dots,p}.$$ 
%This is very convenient when one uses mIPH distributions for survival analysis, given the prevalence of left-truncated data.
%%%%%%%%%%%%%%%%%%%%%%%%%%%%%%%%%%%%%%%%%%%%%%%%%%%%%%%%%%%%%%%%%%
%----------------------------------------------------------------%
%%%%%%%%%%%%%%%%%%%%%%%%%%%%%%%%%%%%%%%%%%%%%%%%%%%%%%%%%%%%%%%%%%
\begin{remark}\normalfont
One might be tempted to argue that Assumption \eqref{dependence_def} necessarily leads to positive dependence of the resulting random variables (in our case lifetimes). However, sharing the initial state is not sufficient to obtain positive dependence, as the different intensity matrices may introduce counter-effects. For instance, after starting in the same state we could have a very small expected holding time and direct absorption for one marginal, while the second has to pass through the entire state space before absorption happens, leading to a very large survival time. This behaviour could very well be reversed when starting in another (but common) state. Consequently, certain combinations of individual intensity matrices may give rise to negative dependence as well.
\end{remark}

\section{Parameter estimation for right-censored data and covariate information}
In the following we firstly introduce the components needed to estimate the parameters of mIPH distributions, when right-censored data is present. Secondly, we present how to estimate initial distribution vectors considering covariate information. Finally, we propose an adapted Expectation Maximisation algorithm, which we name ERMI algorithm. 
\subsection{EM algorithm for right-censored data}
%We provide a fully-explicit EM algorithm for the estimation of a mIPH distribution with covariate dependent initial distribution vectors, when dealing with right-censored absorption times. 
Taking inspiration from Asmussen et al.\ \cite{asmussen(1996)em} and Olsson \cite{olsson(1996)rcens}, we now derive conditional expectations needed in the EM algorithm for mPH distributions, where absorption times are allowed to be right-censored. Since the eventually targeted mIPH distributions are transformed mPH distributions, after transformation of the data the E-Step and M-Step of the algorithm are the same as for the time-homogeneous case.\\
\textcolor{black}{
Let $\mat{X}=(X_1,\dots,X_d)$ be the collection of random variables we are interested in. Let $\mat{x}=\begin{pmatrix}x_{1}^{(m)},\cdots,x_{d}^{(m)}\end{pmatrix}$ be the observations of absorption times assumed to be generated from $\mat{X}\sim\mbox{mPH}(\bfpi,\mathcal{T})$, where $x_i^{(m)}\in\mathbb{R}_+^n$ for $i=1,\dots,d$.
} 
 We assume that the censoring mechanism is independent of the size of the random variables. The marginals $X_i^{(m)}=\min (x_i^{(m)},R_i^{(m)})$ follow  PH$(\bfpi,\mat{T}_i)$ distributions, where $R_i^{(m)}$ is a random censoring point for the $m$-th observation. The realisation of random right-censoring indicators can be found in $\mat{\Delta}=\begin{pmatrix}\delta_{1}^{(m)},\cdots,\delta_{d}^{(m)}\end{pmatrix}$, where elements $\delta_i^{(m)}\in\mathbb{R}_+^n$, $i=1,\dots,d$, are equal to $1$ if the absorption time $x_i^{(m)}$ is fully observed and 0 if $x_i^{(m)}\geq R_i^{(m)}$ is right-censored. 

\textcolor{black}{
The sample $\mat{X}$ is associated to the latent sample paths $\{J_t^{(i,m)}\}_{t\ge0}$, $i=1,\dots,d$, $m=1,\dots,n$, which are not observed. To face this issue, we make the following definitions. Let
\begin{align*}
B_k&=\sum_{i=1}^d\sum_{m=1}^n 1\{J_0^{(i,m)}=k\},\quad k=1,\dots,p,\\
N^{(i)}_{ks}&=\sum_{m=1}^n \sum_{t\ge0}1\{J_{t-}^{(i,m)}=k,J_{t}^{(i,m)}=s\},\quad k,s=1,\dots,p,\:\: i=1,\dots,d,\\
N^{(i)}_{k}&=\sum_{m=1}^n \sum_{t\ge0}1\{J_{t-}^{(i,m)}=k,J_{t}^{(i,m)}=p+1\},\quad k=1,\dots,p,\:\: i=1,\dots,d,\\
Z_k^{(i)}&=\sum_{m=1}^n \int_0^\infty1\{J_t^{(i,m)} =k\}dt, \quad k=1,\dots,p,\:\: i=1,\dots,d.
\end{align*}
$B_k$ is the number of times marginal jump processes start in State $k$, $N^{(i)}_{ks}$ is the number of transitions from State $k$ to $s$ for jump process $i$ and $N^{(i)}_{k}$ is the number of absorptions from State $k$ for jump process $i$. Finally, $Z_k^{(i)}$ is the time spent in state $k$ prior to absorption of jump process $i$. 
These statistics are not observable, but are sufficient to describe the dynamics of the underlying Markov process. Moreover, they are essential to construct an effective EM-like algorithm. 
Then, the completely observed likelihood can be expressed using the sufficient statistics defined above, as
\begin{equation}
\mathcal{L}_c( \bfp , \mathcal{T};\mat{x})=
\left(\prod_{k=1}^{p} {\pi_k}^{B_k}\right) 
\left( \prod_{i=1}^d\prod_{k=1}^{p}\prod_{s\neq k} {t^{(i)}_{ks}}^{N^{(i)}_{ks}}e^{-t^{(i)}_{ks}Z^{(i)}_k}\right) \left(\prod_{k=1}^{p}{t^{(i)}_k}^{N^{(i)}_k}e^{-t^{(i)}_{k}Z^{(i)}_k}\right),   
\end{equation}
which is seen to conveniently fall into the exponential family of distributions, and thus has explicit maximum likelihood estimators.}

With these assumptions, the derivation of $B_k$, $Z_k^{(i)}$, $N_{ks}^{(i)}$ and $N_k^{(i)}$ , for $k,s=1,\dots,p$ and $i=1,\dots,d$, is analogous to the fully uncensored case (see \cite{bladt2022tractable}). The only difference from the fully observed case is that marginals may have right-censored absorption times. To see how this affects the expectation step of the EM algorithm, we give a detailed derivation of $\E(B_k\mid \mat{X}=\mat{x})$.\\
For the $m$-th row of $\bfY$, let $i_{un}^{(m)}$ denote the collection of indices of marginals that are uncensored and similarly $i_{rc}^{(m)}$ for right-censored marginals. Naturally $i_{un}^{(m)}+i_{rc}^{(m)}=d$. Then, the conditional expectation of $B_k$ under right-censoring is
\begin{align*}
 \mathbb{E}(B_k\mid \mat{X}=\mat{x})&=\sum_{i=1}^d\sum_{m=1}^n \mathbb{E}(1\{J_0^{(i,m)}=k\}\mid \mat{X}=\mat{x})\\
&=d\times\sum_{m=1}^n \P(J_0^{(m)}=k\mid \mat{X}=\mat{x})\\
&=d\times\sum_{m=1}^n \frac{\P(J_0^{(m)}=k)\P( X_j\in dx_j^{(m)}, X_l\ge x_l^{(m)}; j\in i_{un}^{(m)}, l\in i_{rc}^{(m)}\mid J_0^{(m)}=k)}{\P( X_j\in dx_j^{(m)}, X_l\ge x_l^{(m)}; j\in i_{un}^{(m)}, l\in i_{rc}^{(m)})}\\
&=d\times\sum_{m=1}^n \frac{\pi_{k}^{(m)} \prod_{j\in i_{un}^{(m)}} {\bfe_k}^{ \mathsf{T}}\exp( \bfT_j x^{(m)}_j) \bft_j \prod_{l\in i_{rc}^{(m)}} {\bfe_k}^{ \mathsf{T}}\exp( \bfT_l x^{(m)}_l) \bfe }{\sum_{s=1}^p \pi_{s}^{(m)} \prod_{j\in i_{un}^{(m)}} {\bfe_s}^{ \mathsf{T}}\exp( \bfT_j x^{(m)}_j) \bft_j \prod_{l\in i_{rc}^{(m)}} {\bfe_s}^{ \mathsf{T}}\exp( \bfT_l x^{(m)}_l) \bfe},
\end{align*}
where we see a mix of marginal densities and survival functions appearing in both the numerator and denominator. This expectation can also be expressed, using the $\mat{\Delta}$ notation, as
\begin{equation*}
     \E(B_k\mid \mat{X}=\mat{x})=d\times \sum_{m=1}^{n}\frac{\pi_{k}^{(m)} \prod_{i=1}^{d}\left(  \ekt  \exp \left(\Ti\xim \right)\ti \right)^{\rc} \left( \ekt  \exp \left(\Ti\xim \right)\evec\right)^{1-\rc} }{ \probObs }
\end{equation*} and we shall use this style in the following. The other needed conditional expectations are obtained in a similar way, reading
\begin{align*}
     \E(Z_{k}^{(i)}\mid \mat{X}=\mat{x})&=\sum_{m=1}^{n}\frac{
    \sum_{j=1}^{p}\pi_{j}^{(m)} \akMinusI
    }{ \probObs}  \times \\
   & \times\left[ \ekt \matInt \ek \right]^{\rc} \times \\
   & \times\left[ \ekt \matIntRc \ek  \right]^{1-\rc},
\end{align*}
\begin{align*}
    \E(N_{ks}^{(i)}\mid \mat{X}=\mat{x})&= t_{ks}^{(i)}\times\sum_{m=1}^{n}\frac{
    \sum_{j=1}^{p}\pi_{j}^{(m)} \akMinusI
    }{ \probObs} \times\\
   & \times\left[\est \matInt\ek \right]^{\rc} \times \\
    &\times \left[\est \matIntRc \ek \right]^{1-\rc},
    \end{align*}
and finally   
\begin{align*}
    \E(N_{k}^{(i)}\mid \mat{X}=\mat{x})&=t_{k}^{(i)} \times \sum_{m=1}^{n} \sum_{j=1}^{p} \pi_{j}^{(m)} \ejt \exp(\Ti \xim)\ek \rc \times\\
    & \times \frac{
    \akMinusI }{ \probObs}.
\end{align*}
\subsection{Initial distribution vectors}
Adapting an idea developed in \cite{BladtYslas2022}, we apply the regression component to the initial distribution, i.e.\ we estimate a `personalised' initial distribution vector as a function of covariates, which increases the flexibility of the model. To that end, we use multinomial logistic regressions, where we consider the initial probabilities as response variables that depend on covariate information found in ${\mat{A}^{(m)}}^{\mathsf{T}} \in \mathbb{R}^{g}$, with $g$ being the number of explanatory variables,  $m=1,\dots,n$, and regression coefficients in $\vect{\gamma}\in \mathbb{R}^{p\times g}$. Concretely, the initial distribution probabilities are then given as
$$
\pi^{(m)}_k=\frac{\exp{\left({\mat{A}^{(m)}}\gamma_k\right)}}{\sum_{j=1}^{p}\exp{\left({\mat{A}^{(m)}}\gamma_j\right)}},
$$
with $\gamma_k\in\mathbb{R}^g$ for $k=1,\dots,p$.\\
In every iteration of the expectation-maximisation (EM) algorithm to be described later, we use the conditional expectation of the number of times that the underlying process starts in a specific state as weights for the regression coefficients in $\vect{\gamma}$. 
Let us consider the information carried by $\E(B_k\mid \mat{X}=\mat{x})$ separately. For a row of observations $m$, the expected number of times that the marginal jump processes start in state $k$ is
\[
 \mathbb{E}(B_{k}^{(m)}\mid \mat{X}=\mat{x})=d\times \frac{\pi_{k}^{(m)} \prod_{i=1}^{d}\left(  \ekt  \exp \left(\Ti\xim \right)\ti \right)^{\rc} \left( \ekt  \exp \left(\Ti\xim \right)\evec\right)^{1-\rc} }{ \probObs },
\] 
for $k=1,\dots,p$. We then solve the optimisation problem
$$
 \hat{ \vect{\gamma}}= \argmax_{\vect{\gamma}} \sum_{m=1}^{n} \sum_{k=1}^{p} \E(B_{k}^{(m)}\mid \mat{X}=\mat{x}) \log(\pi_{k}^{(m)}(\mat{A}^{(m)};\vect{\gamma}))
$$ and set
\begin{align}
    \hat \pi_k^{(m)}=\pi_k^{(m)}(\mat{A}^{(m)}; \hat{\vect{\gamma}})=\frac{\exp{\left({\mat{A}^{(m)}}\hat \gamma_k\right)}}{\sum_{j=1}^{p}\exp{\left({\mat{A}^{(m)}}\hat\gamma_j\right)}}\label{tt}
\end{align}
in every iteration. 
%This allows components of a random vector to share different starting probabilities for the same state-space, increasing the precision of the estimation itself.
The initial distribution hence depends on covariate information. Recall that all marginal processes $\{J_t^{(i)}\}_{t\geq0}$ are assumed to start in the same state (drawn from the initial distribution with probabilities \eqref{tt}), but afterwards transit independently to other states according to their specific sub-intensity matrices (and the latter do not depend on covariate information). 
%Their estimation is described next.
\subsection{ERMI algorithm.}
Consider now a multivariate sample of right-censored absorption times $\mat{y}=\begin{pmatrix}y_{1}^{(m)},\cdots,y_{d}^{(m)}\end{pmatrix}$, which we assume to originate from $Y^{(m)}\sim \mbox{mIPH}(\bfp^{(m)},\mathcal{T},\mathcal{L})$. The associated inhomogeneity functions depend on parameters $\beta_i$, $i=1,\dots,d$, and the right-censoring indicators are collected in $\mat{\Delta}$. The resulting EM algorithm with covariate information is depicted in Algorithm \ref{alg1}.
%EM algorithm
\begin{algorithm}[t]
	\caption{\textit{Adapted expectation maximisation (ERMI) algorithm for mIPH distributions}}
	\label{alg1} 
	\begin{algorithmic}
		\State \textit{\textbf{Input}: Observed absorption times $\mat{y}\in\mathbb{R}^{n\times d}_+$, right-censoring indicators $\mat{\Delta}\in\mathbb{R}^{n\times d}$ and arbitrary initial parameters for $( \bfpi, \calT, \calL)$.}\\
		\begin{enumerate} 
			\item[ 1)]\textit{For each marginal, transform the data in $x_{i}^{(m)}=g_{i}^{-1}(y_{i}^{(m)};\beta_i)$, $i=1,2,\dots,d$ and $m=1,2,\dots,n$}
			\item[ 2)]\textit{E-step: Calculate} 
			\begin{align*}
				&\E(B_{k}^{(m)} \mid \mat{Y}=\mat{y})\quad k=1,\dots,p\,, \;\; m=1,\dots,n\\
				&\E(Z_k^{(i)} \mid \mat{Y}=\mat{y})\quad k=1,\dots,p\,, \;\; i=1,\dots,d\\
				&\E(N_{ks}^{(i)} \mid \mat{Y}=\mat{y})\quad k,s=1,\dots,p\,, \;\; i=1,\dots,d\\
				&\E(N_k^{(i)} \mid \mat{Y}=\mat{y})\quad k=1,\dots,p\,, \;\; i=1,\dots,d
			\end{align*}
			\item[3)]\textit{ R-step: Perform a multinomial regression with weights given by $\E(B_{k}^{(m)}\mid \mat{Y}=\mat{y})$ and predict $\hat \pi_{k}^{(m)}$ for $k=1,\dots,p$ and $m=1,\dots,n$.}
			\item[4)] \textit{M-step: Let} 
			\begin{align*}
				%\hat \pi_{k}^{(m)}&=\frac{\mathbb{E}(B_k\mid \mat{Y}=\mat{y})}{n\cdot d}, \quad k=1,\dots,p\\
				\hat t_{ks}^{(i)}&=\frac{\mathbb{E}(N_{ks}^{(i)}\mid \mat{Y}=\mat{y})}{\mathbb{E}(Z_{k}^{(i)}\mid \mat{Y}=\mat{y})}, \quad k,s=1,\dots,p\,, \;\; i=1,\dots,d\\
				\hat t_{k}^{(i)}&=\frac{\mathbb{E}(N_k^{(i)}\mid \mat{Y}=\mat{y})}{\mathbb{E}(Z_{k}^{(i)}\mid \mat{Y}=\mat{y})} \quad k=1,\dots,p\,, \;\; i=1,\dots,d\\
				\hat t_{kk}^{(i)}&=-\sum_{s\neq k}\hat t_{ks}^{(i)}-\hat t_k^{(i)}\quad k,s=1,\dots,p\,, \;\; i=1,\dots,d
			\end{align*}
			Let $\hat\bfpi=( \hat{\pi}_{1}^{(m)}, \cdots ,\: \hat{\pi}_{p}^{(m)}),\quad \hat{\bfT_i} =\{{ \hat{t}}_{ks}^{(i)}\}_{k,s=1,2,\cdots,p},\quad \mbox{and} \quad \hat{\bft_i}=\begin{pmatrix} \hat{t}_1^{(i)} \\ \vdots \\ \hat{t}_p^{(i)} \end{pmatrix}$.
			\item[5)] \textit{ I-step: Compute}
			\begin{align*}
				\hat \bfBeta &= \argmax_{\bfBeta} \sum_{m=1}^{n} \log \left(f_{Y}(\mat{y}^{(m)};\hat \bfpi,\hat \calT,\bfBeta,\mat{\Delta})\right)\\
				&=\argmax_{\bfBeta} \sum_{m=1}^{n} \log \left( \sum_{j=1}^p\hat\pi_{j}^{(m)} \prod_{i=1}^d
				\left(\bfe_j^\mathsf{T}\exp(\hat \bfT_i x_i^{(m)})\hat \bft_i\lambda_i(y_i^{(m)},\beta_i)\right)^{\delta_i^{(m)}}
				\left(\bfe_j^\mathsf{T}\exp(\hat \bfT_i x_i^{(m)})\bfe \right)^{1-\delta_i^{(m)}}
				\right)
			\end{align*}
			\item[6)] \textit{Assign $\bfpi =\hat\bfpi$, $\bfT_i=\hat{\bfT_i}$ and $\beta_i=\hat \beta_i$ then repeat from Step 1 until a stopping rule is satisfied.}
		\end{enumerate}
		\State \textit{\textbf{Output}: Fitted representations $(\hat \bfpi ,\hat \calT,\hat \calL )$, for $m=1,\dots,n$.}
	\end{algorithmic}
\end{algorithm}
As in Albrecher et al.\ \cite{ABY22}, we firstly take care of time-inhomogeneity. Using the relation $g^{-1}_i\left(y_i^{(m)}\right)=\int_{0}^{y_i^{(m)}}\lambda_i(u;\beta_i)du,\quad i=1,\dots, d,$ we obtain a time-homogeneous random sample $\left(x_1^{(m)},\dots,x_d^{(m)}\right)$, for which we know how to evaluate conditional expectations of sufficient statistics (E step). Using these expectations (as given above), we estimate the marginal sub-intensity matrices (M step), while the initial distribution vectors are predicted by multinomial logistic regressions (R step). Once we have estimated both, we need to find optimal inhomogeneity parameters $\beta_i$, $i=1,\dots,d$, that maximise the joint likelihood of the time-inhomogeneous sample (I step). Concretely, we solve
\begin{align*}
    \hat \bfBeta &=\argmax_{\bfBeta} \sum_{m=1}^{n} \log \left( \sum_{j=1}^p\hat\pi_{j}^{(m)} \prod_{i=1}^d
    \left(\bfe_j^\mathsf{T}\exp(\hat \bfT_i x_i^{(m)})\hat \bft_i\lambda_i(y_i^{(m)},\beta_i)\right)^{\delta_i^{(m)}}
    \left(\bfe_j^\mathsf{T}\exp(\hat \bfT_i x_i^{(m)})\bfe \right)^{1-\delta_i^{(m)}}
 \right),
\end{align*}where $ x_i^{(m)}= g_i^{-1}(y_i^{(m)})$. We repeat the procedure until a stopping rule is satisfied, and finally obtain the estimated distribution mIPH$(\hat \bfpi ,\hat \calT,\hat \calL )$. Here $\hat \bfpi$ is a matrix, where each row is a distribution vector $\hat \bfpi^{(m)}$, which is shared by marginals with the same covariates.

\textcolor{black}{
In contrast to copula-based methods, this approach does not separate the estimation of marginals and multivariate parameters. This may be considered preferable as the implied  multivariate distribution has a natural and causal interpretation and is intimately connected to the marginal behaviour of the risks, whereas choosing a concrete copula family on given (and possibly already fitted) marginal risks is often a somewhat more arbitrary choice for the modelling of multivariate phenomena.   
%
%since marginals are the only responsible for the dependence structure. The denseness property embedded in the mIPH class allows for very precise estimation of both marginal and multivariate distributions. 
%Thus, instead of assuming a particular copula family, dependence between marginals is modelled directly by the representation $(\bfp,\mathcal{T},\mathcal{L})$. As a consequence, one avoids making additional assumptions about the interaction between marginals, e.g. choosing a parametric copula family.   
}
%With a mIPH distribution optimal parameters of marginal distributions are directly considered in the description of multivariate random vectors. 
%Instead of an "$\alpha$'' parameter, the shared initial distribution vector and marginals characteristics influence together the dependence of marginal components.      
%%%%%%%%%%%%%%%%%%%%%%%%%%%%%%%%%%%%%%%%%%%%%%%%%%%%%%%%%%%%%%%%%%
%----------------------------------------------------------------%
%%%%%%%%%%%%%%%%%%%%%%%%%%%%%%%%%%%%%%%%%%%%%%%%%%%%%%%%%%%%%%%%%%
\section{Modelling joint excess lifetimes of couples}
In this section we present an application of Algorithm \ref{alg1} to the well-known data set of joint lives used in \cite{frees(1996)}. In addition to survival times, this data set provides information on individuals' ages at issue of an insurance policy.
This leads to left-truncated data in this particular case, and one might indeed consider the different entry ages with a left-truncated likelihood in the estimation process, which is, however, quite inefficient. Instead, we propose here to use entry age as a covariate information and use multinomial logistic regression to deal with the different ages at issue. Entry ages were also considered as relevant factors for the dependence modelling in \cite{shemyakin2006}. 
%In order to have a precise approximation via the EM algorithm, we need multiple data points which we assume to generate from the same distribution. To include the left-truncated likelihood we would need many couples with exactly the same (or very similar) ages at issue of the policy. If it was the case, the estimation procedure illustrated above could be employed to produce mIPH distributions for each age combinations of couples we see in the data. However, although we do not fully consider left-truncation we think that the procedure we are about to describe is innovative and worthwhile of consideration, especially in a large data environment.\\
%We want the resulting multivariate distributions to reflect the age of the individuals in a couple. 
Initial distribution vectors obtained via regression will then incorporate the fact that an old couple is expected to survive less long than a young one, and that the bereavement effect may be different for different age dynamics in a couple. In PH terms, when using an acyclical distribution, the older the couple the larger the starting probabilities should be for states closer to the absorbing state, so that fewer states will be visited until absorption.  Passing through fewer states translates into less time spent in the state-space before exiting, which results in smaller remaining lifetimes.
\subsection{Description of the data set}
The data set at hand provides information about 14947 insurance products on joint lives, which were observed from December $29$, $1988$ until December $31$, $1993$. We consider January 1, 1994 as the right censoring limit.
For the purpose of this paper, we only consider birthdays, sexes (and potential death dates) of policyholders, given that we are only interested in mortality, i.e.\ we do not make use of the monetary details of each contract. \\
After removing same-sex couples and multiple entries (due to several contracts of the same couple), we compute the remaining lifetime that any person lived from the start of the observation period until the right-censoring date $01.01.1994$, given that they are at least $40$ years old at the issue of the policy. 
\textcolor{black}{Note that the terms ``remaining" and ``excess" are used interchangeably in the following}. 
Doing so leads to $8834$ different joint excess survival times, with $155$ cases where both individuals died, $1057$ where only one individual died and $7622$ where neither died.  Consequently, less than $2\%$ of joint remaining survival times are fully observed. Among the couples where only one person died, in $820$ of them this was the man, and in the complementary $237$ cases the woman died. Hereafter, we refer to the start of the observation period as the ``issue of the policy", although the actual issue date may be later than $29.12.1988$.\\
%To prepare data for the multinomial logistic regression, we partition all couples into $4$ subgroups. In the first we find all couples where both partners were younger than $\xx$ at issue. The second and third have data for couples where one partner was younger than $\xx$ and the other older, with Subgroup 2 being the one where the man was younger than $\xx$ and in Subgroup 3 the woman was younger than $\xx$. Finally, Subgroup 4 contains all information on couples where both individuals were older than $\xx$ at the issue of respective policies. %Old version
To prepare the data for the multinomial regression, we construct the covariate matrix $\mat{A}=\begin{pmatrix} \vect{1} & age_{y_1} & age_{y_2} & interact \end{pmatrix}$, $\mat{A}\in \mathbb{R}^{n\times4} $. The column vector $age_{y_1}$ is the collection of all ages of men, at issue of a policy, while $age_{y_2}$ contains all ages of women. We also consider an interaction term in $interact$, which gathers element-wise multiplication of ages in a couple. Finally, to perform multinomial regressions via neural networks (R package \textit{nnet}), we divide the data by $100$, such that an absorption time of $0.01$ given by estimated distributions actually corresponds to $1$ year. 
%%%%%%%%%%%%%%%%%%%%%%%%%%%%%%%%%%%%%%%%%%%%%%%%%%%%%%%%%%%%%%%%%%%%%%%%%%%%%%%%%%%%%%%%%%%%
\subsection{Fitting the mIPH distribution: Marginal behaviour}%{Optimal representations}
We assume that the remaining lifetimes in a couple, after issue of an insurance policy, follow mIPH distributions. We use information previously discussed as covariates to link starting probabilities to ages of individuals in couples, in order to reflect different ageing dynamics in our model. 

Let $Y_{i}^{(m)}=g_i(X_i^{(m)})=\log(\beta_i X_{i}^{(m)}+1)/\beta_i$, be the IPH distributed marginal remaining lifetimes, with $X_{i}^{(m)}=\left(\exp\left(\beta_iY_i^{(m)}\right)-1\right)/\beta_i\sim \mbox{PH}(\bfpi^{(m)},\Ti)$ and $\beta_i > 0$, where $i=1$ corresponds to men and $i=2$ to women. Then, the distribution with covariate information we work with is given by marginal Matrix-Gompertz distributions, and we consider general Coxian sub-intensity matrices $\Ti$ of dimension $p=10$ (cf.\ \cite{ABBY21}).
\textcolor{black}{
The order $p$ was found by fitting models with various respective choices, until a satisfactory approximation of marginals was reached.
Finding the optimal dimension $p$ is still an open problem, and the literature on solving the identifiability issue of PH distributions is quite narrow (see for instance \cite{faddy2002penalised} and \cite{albrecher2022penalised}).}

According to the dependence structure defined in \eqref{dependence_def}, after both underlying processes $\{ J_t^{(1)} \}_{t \geq 0}$ and $\{ J_t^{(2)} \}_{t \geq 0}$ start in the same state they evolve independently until absorption. With a general Coxian sub-intensity matrix, each marginal jump process is only allowed to transit to the next state or directly to the absorption state. This stochastic structure has a nice interpretation in terms of ageing. Indeed, we can think of forward transitions as natural ageing steps, given that each time the process jumps to a certain state, the time of absorption gets closer. Moreover, premature exits can be interpreted as deaths due to causes not related to ageing. Indeed, given that exit rates are positive in each state, the absorption of a process may be caused by a transition to State $p+1$ from a state smaller than $p$. Finally, granting the underlying processes to start in different states allows heterogeneity of health statuses for individuals of the same age (this philosophy was already underlying the construction in \cite{markovageing2007Lin&Liu}, but with the present inhomogeneity, much fewer states are needed to describe the data satisfactorily).

\begin{figure}[ht]
    \centering
    \includegraphics[width=0.6\textwidth]{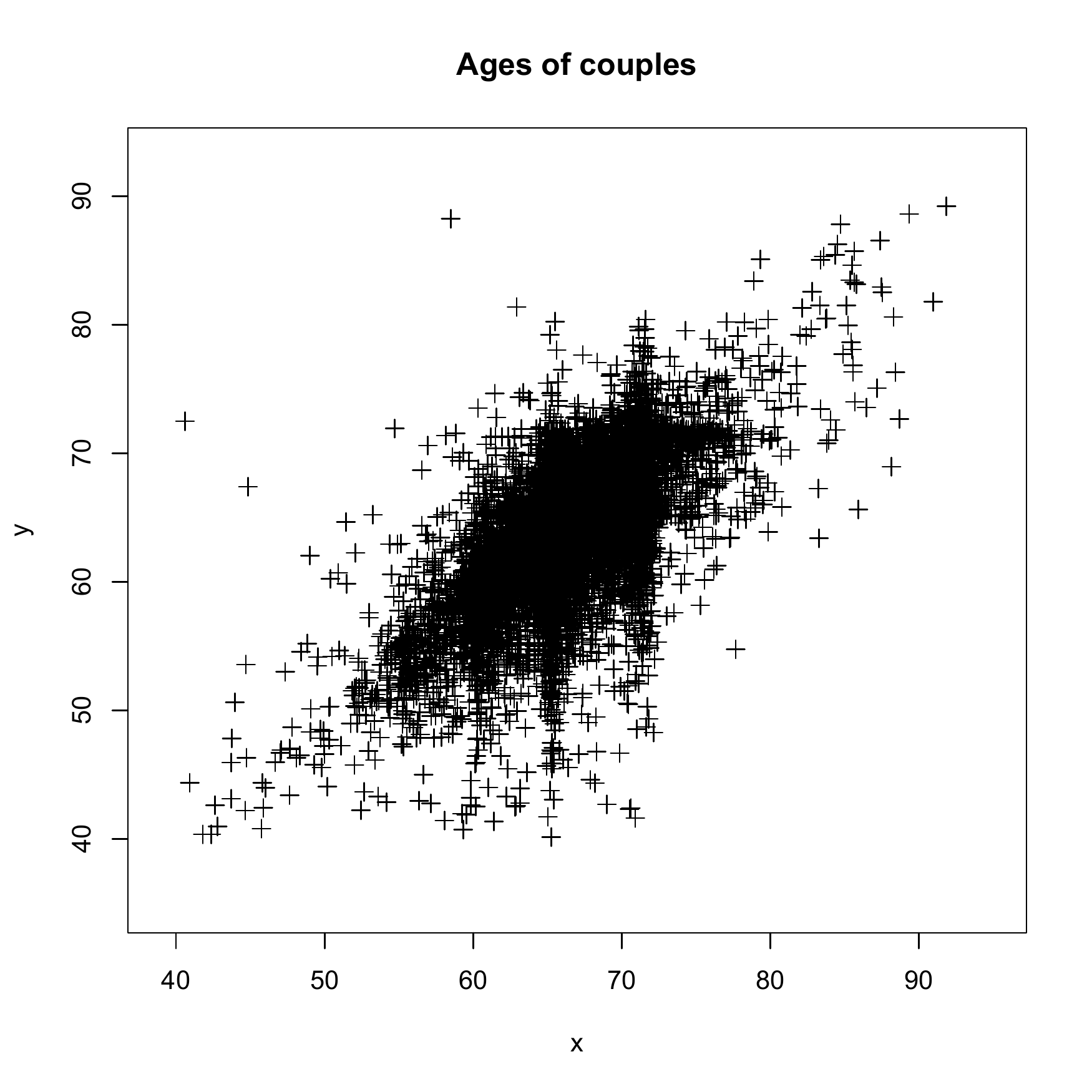}
    \caption{Ages at issue of policies, for all couples. $x$ and $y$ correspond to $age_{y_1}$ and $age_{y_1}$, respectively.}
    \label{fig: age coord}
\end{figure}

Algorithm \ref{alg1} is now applied for $1000$ iterations.
%As a consequence of not considering left-truncated likelihoods in the estimation process, inference on total lifetimes of couples is not possible. The distinct distributions we obtain only model joint excess survival times, after issue of a policy. They do not consider the actual age of individuals as starting ages to which one would add the excess. Differences between men and women are modelled bivariately, by virtue of gender specific sub-intensity matrices and inhomogeneity functions.\\
In principle, this estimation procedure provides $n=8834$ different initial distribution vectors, one for each couple depending on the ages at issue. Figure \ref{fig: age coord} depicts the age combinations in the data, with a majority of couples having a small age  difference at policy issue. 
For illustration purposes, we depict below the estimated initial distributions for four different age combinations (age man, age woman): ($63$,$63$), ($68$,$63$), ($63$,$68$) and ($73$,$63$), which were chosen arbitrarily, but to represent different ageing dynamics at issue of the policy. 
%
%In the first couple we consider an age difference of less than one year. For the second and third couple we have age differences of five years, the woman and the man being younger, respectively. Lastly, in the fourth couple the man is ten years older.\\ 
Let $\mat{Y}^c=(Y_1^c,Y_2^c)\sim\mbox{mIPH}(\hat \bfp^c,\hat \calT,\hat \calL)$ denote the bivariate distribution of excess lifetimes for these four couples, where $c=1,2,3,4$, $\hat \calT=\{\hat \bfT_1,\hat \bfT_2\}$ and $\hat \calL=\{\lambda_1(\cdot,\hat \beta_1),\lambda_2(\cdot,\hat \beta_2)\}$. The common estimated sex-specific sub-intensity matrices are
{ \begin{align}\label{c1 T1}
 \hat\bfT_1=\begin{pmatrix} 
-0.049 & 1.7/10^{7} & 0 & 0 & 0 & 0 & 0 & 0 & 0 & 0 \\ 
0 & -3.662 & 2.877 & 0 & 0 & 0 & 0 & 0 & 0 & 0 \\ 
0 & 0 & -1.8/10^{7} & 1.8/10^{7} & 0 & 0 & 0 & 0 & 0 & 0 \\ 
0 & 0 & 0 & -1.9/10^{4} & 1.9/10^{4} & 0 & 0 & 0 & 0 & 0 \\ 
0 & 0 & 0 & 0 & -0.611 & 0.611 & 0 & 0 & 0 & 0 \\ 
0 & 0 & 0 & 0 & 0 & -0.002 & 0.002 & 0 & 0 & 0 \\ 
0 & 0 & 0 & 0 & 0 & 0 & -9.778 & 5.73 & 0 & 0 \\ 
0 & 0 & 0 & 0 & 0 & 0 & 0 & -0.36 & 0.225 & 0 \\ 
0 & 0 & 0 & 0 & 0 & 0 & 0 & 0 & -1.852 & 1.099 \\ 
0 & 0 & 0 & 0 & 0 & 0 & 0 & 0 & 0 & -0.023 
\end{pmatrix}   
\end{align}}
for men and 
{ \begin{align}\label{c1 T2}
\hat \bfT_2=\begin{pmatrix} 
-0.196 & 0.196 & 0 & 0 & 0 & 0 & 0 & 0 & 0 & 0 \\ 
0 & -0.291 & 0.291 & 0 & 0 & 0 & 0 & 0 & 0 & 0 \\ 
0 & 0 & -0.763 & 0.763 & 0 & 0 & 0 & 0 & 0 & 0 \\ 
0 & 0 & 0 & -2.8/10^{8} & 2.8/10^{8} & 0 & 0 & 0 & 0 & 0 \\ 
0 & 0 & 0 & 0 & -0.001 & 0.001 & 0 & 0 & 0 & 0 \\ 
0 & 0 & 0 & 0 & 0 & -0.003 & 0.003 & 0 & 0 & 0 \\ 
0 & 0 & 0 & 0 & 0 & 0 & -3.182 & 1.165 & 0 & 0 \\ 
0 & 0 & 0 & 0 & 0 & 0 & 0 & -0.172 & 2/10^{7} & 0 \\ 
0 & 0 & 0 & 0 & 0 & 0 & 0 & 0 & -0.008 & 2.3/10^{10} \\ 
0 & 0 & 0 & 0 & 0 & 0 & 0 & 0 & 0 & -3/10^{6}  
\end{pmatrix}   
\end{align}}
for women. The optimal inhomogeneity parameters are $\hat \beta_1=43.101$ and $\hat \beta_2=47.474$, and Table \ref{table: coef} presents the estimated coefficients of the last multinomial regression performed in Algorithm \ref{alg1}.
\begin{table}[ht]
\centering
\begin{tabular}[t]{ccccc}
\toprule
$p$ & Intercept & $age_{y_1}$ & $age_{y_2}$ & $age_{y_1} \cdot age_{y_2}$\\
\midrule
2 & $-20.963^{***}$ & $43.733^{***}$ & $43.021^{***}$ & $-84.049^{***}$\\
\addlinespace
 & (7.157) & (11.238) & (12.400) & (19.311)\\
\addlinespace
3 & $24.826^{***}$ & $-24.630^{**}$ & $-39.256^{***}$ & $38.453^{**}$\\
\addlinespace
 & (6.677) & (10.476) & (11.639) & (18.073)\\
\addlinespace
4 & $-51.036^{***}$ & $57.442^{***}$ & $90.062^{***}$ & $-104.233^{***}$\\
\addlinespace
 & (9.894) & (15.348) & (15.820) & (24.374)\\
\addlinespace
5 & $-42.469^{***}$ & $56.804^{***}$ & $70.273^{***}$ & $-89.556^{***}$\\
\addlinespace
 & (6.873) & (10.638) & (11.443) & (17.559)\\
\addlinespace
6 & $14.850^{*}$ & $-41.377^{***}$ & $-39.438^{***}$ & $89.687^{***}$\\
\addlinespace
 & (8.004) & (12.150) & (12.824) & (19.317)\\
\addlinespace
7 & $54.157^{***}$ & $-97.618^{***}$ & $-98.445^{***}$ & $173.553^{***}$\\
\addlinespace
 & (6.617) & (10.263) & (11.010) & (16.817)\\
\addlinespace
8 & -14.363 & -5.608 & 22.990 & 8.794\\
\addlinespace
 & (9.419) & (14.387) & (14.885) & (22.584)\\
\addlinespace
9 & -11.589 & 12.732 & -4.892 & 18.559\\
\addlinespace
 & (7.224) & (11.046) & (11.889) & (18.050)\\
\addlinespace
10 & $21.474^{***}$ & $-31.068^{***}$ & $-54.907^{***}$ & $84.080^{***}$\\
\addlinespace
 & (6.504) & (10.054) & (10.892) & (16.670)\\
\bottomrule
\end{tabular}
\caption{Coefficients of multinomial regression with associated standard errors in brackets. For a two sided statistical test symbols $***$, $**$ and $*$ correspond to significance levels of $1\%$, $5\%$ and $10\%$ respectively.}\label{table: coef}
\end{table}
Except for States 8 and 9, all coefficients in $\hat{\mat{\gamma}}$ are significant.
\textcolor{black}{
Note that having non-significant coefficients for States 8 and 9 does not necessarily indicate that the state space is too large. One has to separate the PH interpretation of the state space from the regression framework. Table \ref{table: coef} is claiming that the covariates used in the regression are not significant in explaining starting probabilities of States 8 and 9. Nevertheless, both states have their role in the description of random variables we are interested in.}

Initial distribution vector estimates for these four couples are 
\begin{align*}
\hat \bfp^1 &=\begin{pmatrix} 
0.0526 & 0.0734 & 0.0448 & 0.0886 & 0.4065 & 0.0330 & 0.0326 & 0.0569 & 0.1077 & 0.1039 \end{pmatrix},\\
\hat \bfp^2&=\begin{pmatrix} 
0.0356 & 0.0313 & 0.0297 & 0.0398 & 0.2805 & 0.0476 & 0.0396 & 0.0384 & 0.2472 & 0.2102
\end{pmatrix},\\
\hat \bfp^3&=\begin{pmatrix} 
0.0285 & 0.0242 & 0.0114 & 0.1625 & 0.4399 & 0.0419 & 0.0304 & 0.1282 & 0.0819 & 0.0510  \end{pmatrix},\\
\hat \bfp^4&=\begin{pmatrix} 
0.0172 & 0.0095 & 0.0140 & 0.0127 & 0.1378 & 0.0489 & 0.0343 & 0.0184 & 0.4041 & 0.3030 \end{pmatrix}.
\end{align*}

At first glance, it might seem odd that survival times of spouses with large age difference start in the same state of the distribution, but personalised starting probabilities and sex-specific transition intensities account for this. For example, consider Couple 4. If both excess survival times start in State 1, we see that the man's underlying jump process is much more likely to reach the absorbing state directly from State 1, while the woman's will at least advance to State 7 before absorption becomes possible. Thus, marginal intensities mixed with age-dependent initial distribution vectors compensate the initialisation in a shared state.
\begin{figure}[ht]
    \centering
    \begin{subfigure}[b]{0.495\textwidth}
        \includegraphics[width=1\textwidth]{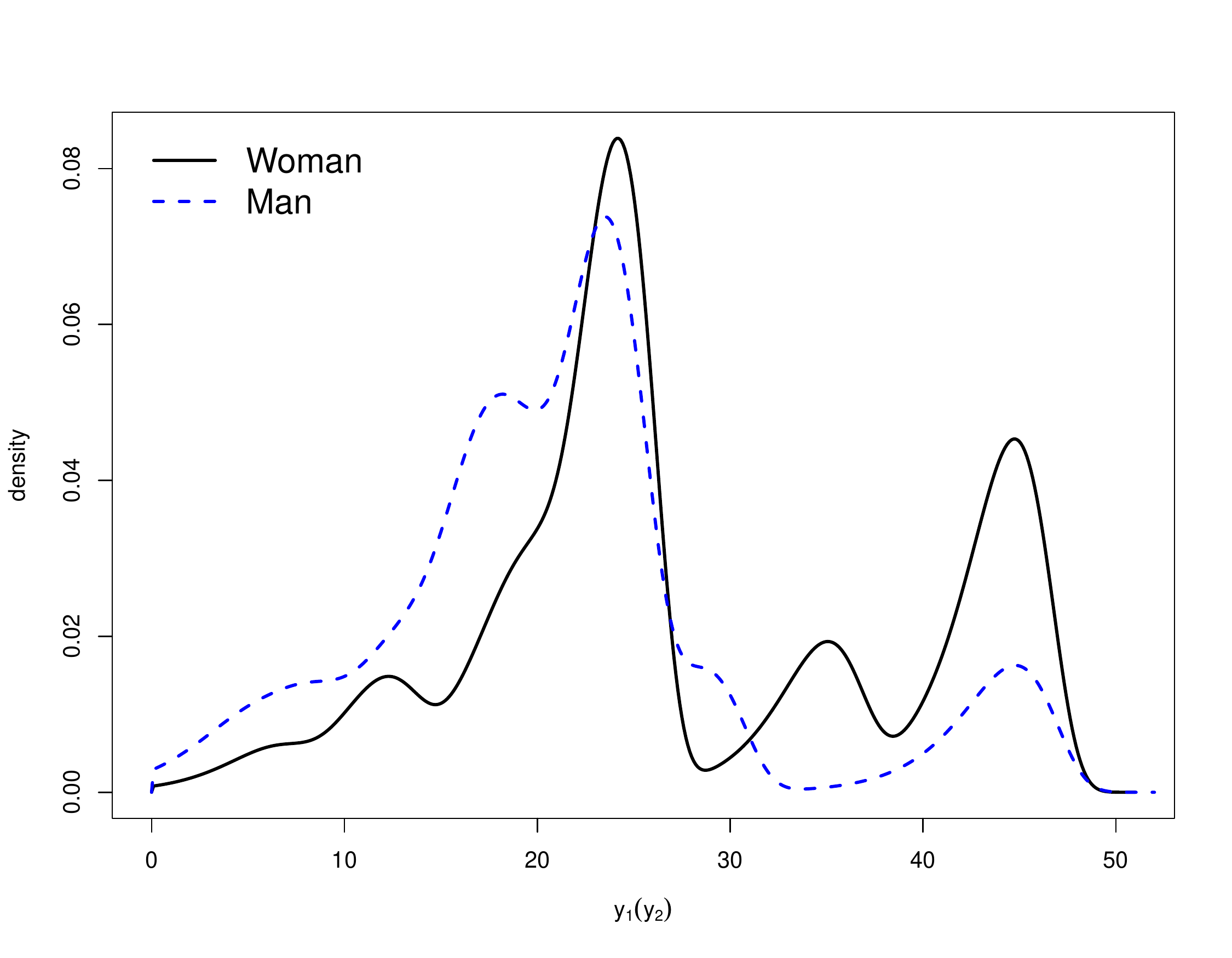}
        \caption{Couple 1}
    \end{subfigure}
    \hfill
        \begin{subfigure}[b]{0.495\textwidth}
        \includegraphics[width=1\textwidth]{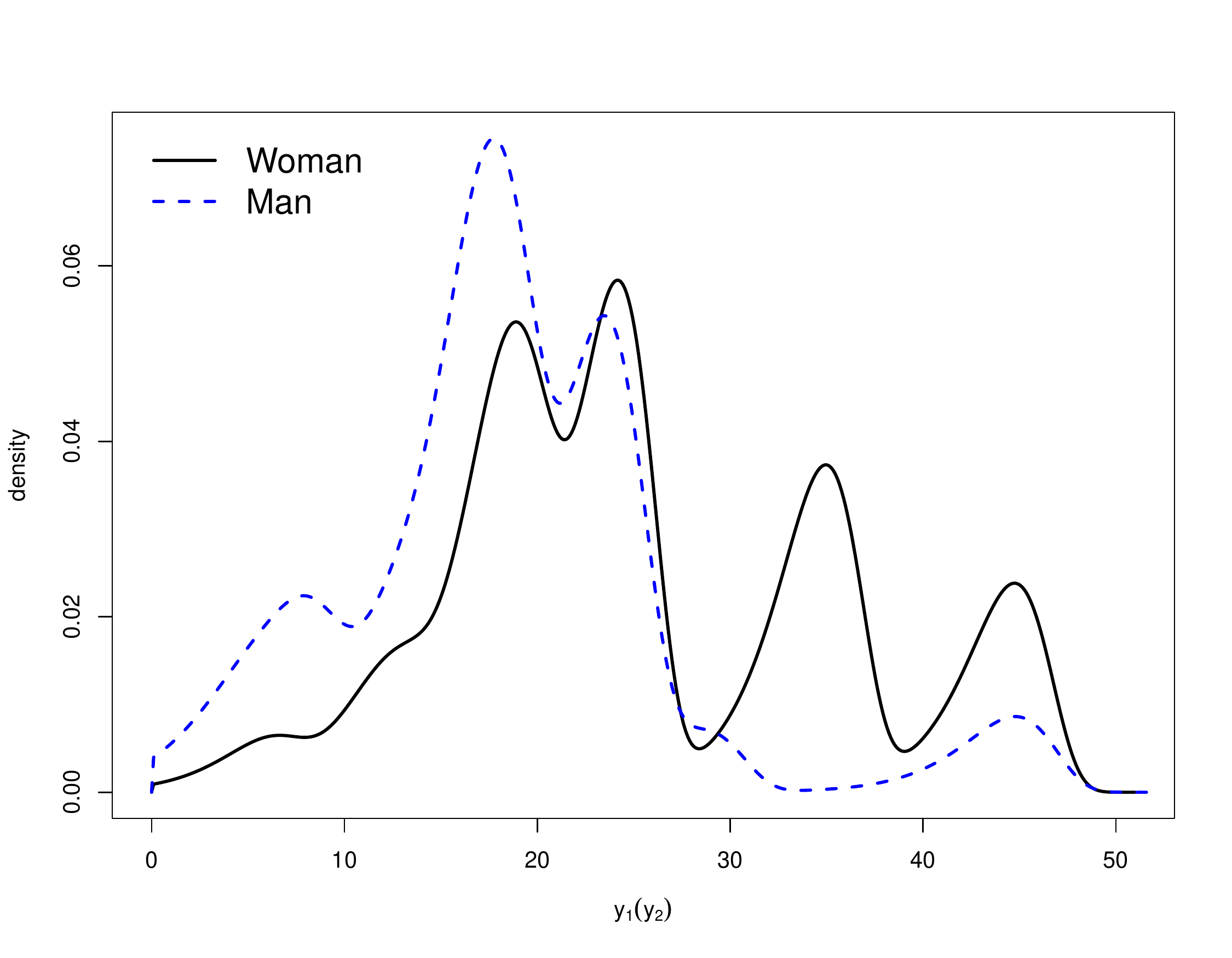}
        \caption{Couple 2}
    \end{subfigure}
    \vfill
        \centering
    \begin{subfigure}[b]{0.495\textwidth}
        \includegraphics[width=1\textwidth]{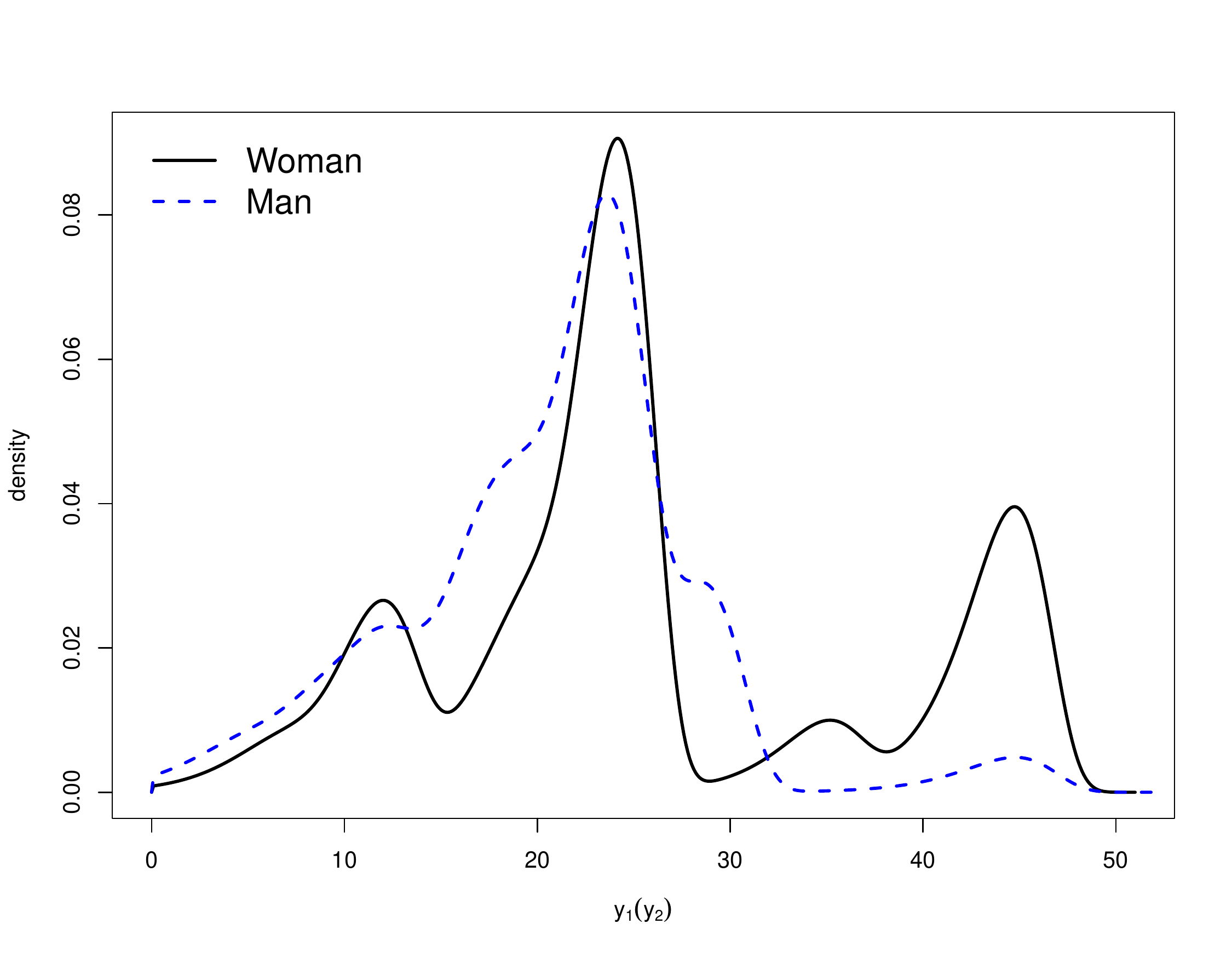}
        \caption{Couple 3}
    \end{subfigure}
    \hfill
        \begin{subfigure}[b]{0.495\textwidth}
        \includegraphics[width=1\textwidth]{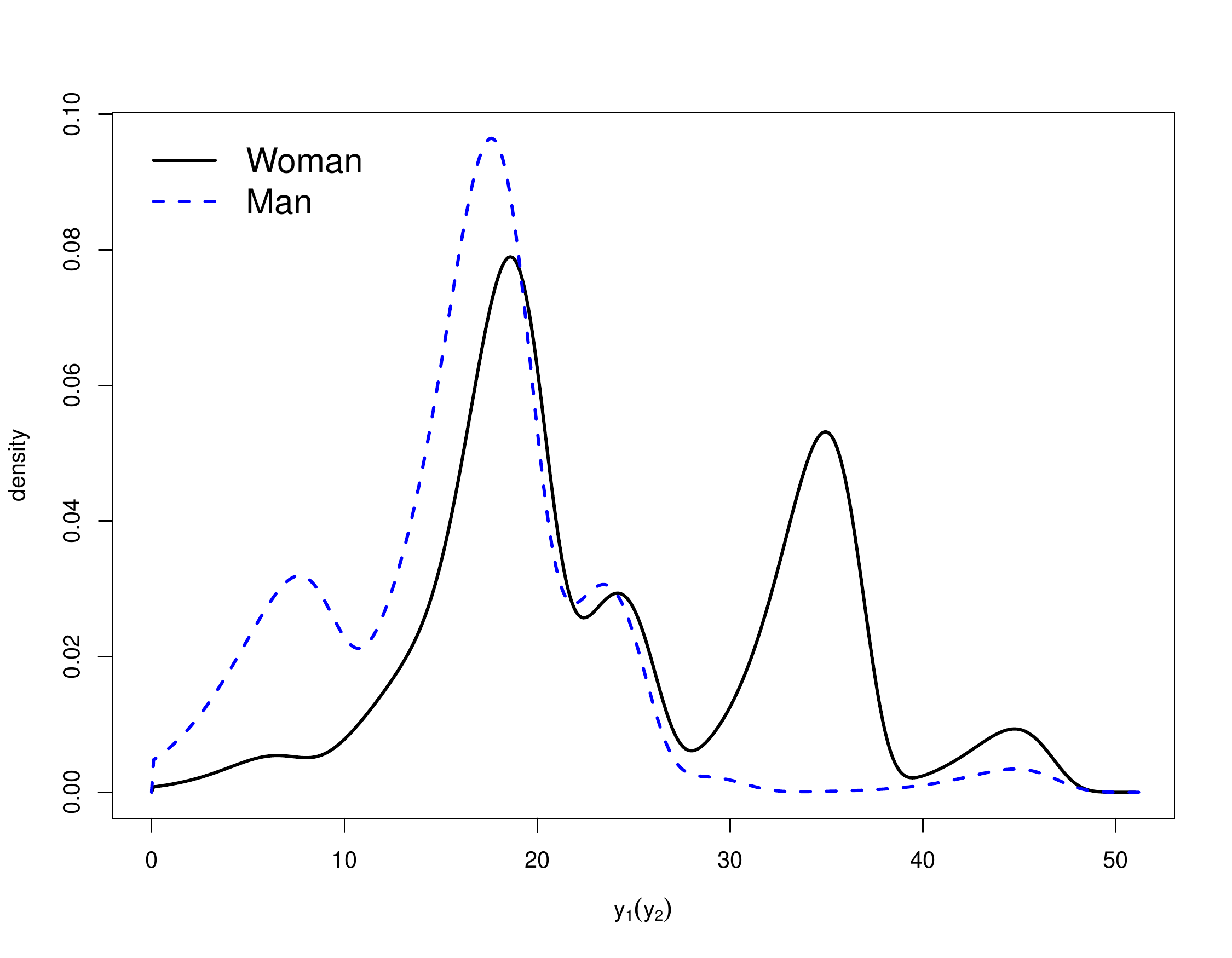}
        \caption{Couple 4}
    \end{subfigure}
    \caption{Marginal densities of remaining life times (in years).}
    \label{fig: pdf comp}
\end{figure}
In Figure \ref{fig: pdf comp} we depict the resulting marginal densities of the fitted remaining lifetime distributions for each of the four couples. As expected, the densities for women (solid black lines) allocate more mass to larger values than the male counterpart, and for older individuals there is more probability mass for shorter survival times. Comparing Couples 1 and 4, we see that the density of the $63$-year-old man has a major mode at $y_1=22$, while for the $73$-year-old man, the major mode is at $y_1=17$. Note that despite the fact that the women in Couples 1 and 4 are of comparable age, their densities have substantially different modes. This is due to the fact that the estimation of the marginal distributions is not separated from the estimation of the joint distributions, and the age of their spouse at the time of policy issue evidently plays an important role for the distribution of the remaining lifetime, seen at the time of policy issue.

The multimodality we observe in all marginal densities may be due to having different cohorts in the data. We decided not to manipulate the data set further, in order to avoid restricting our analysis to specific couples. Although spouses with small age difference can be thought of as belonging to the same cohort, we would also need to restrain the ages at issue to instances where enough data points are available for a meaningful estimation procedure (in particular uncensored data points). Doing so would lead to analysing only couples where both spouses are aged around $65$ years old.
%, couples with other age dynamic would be disregarded. 

\begin{figure}[ht]
    \centering
    \begin{subfigure}[b]{0.9\textwidth}
        \includegraphics[width=1\textwidth]{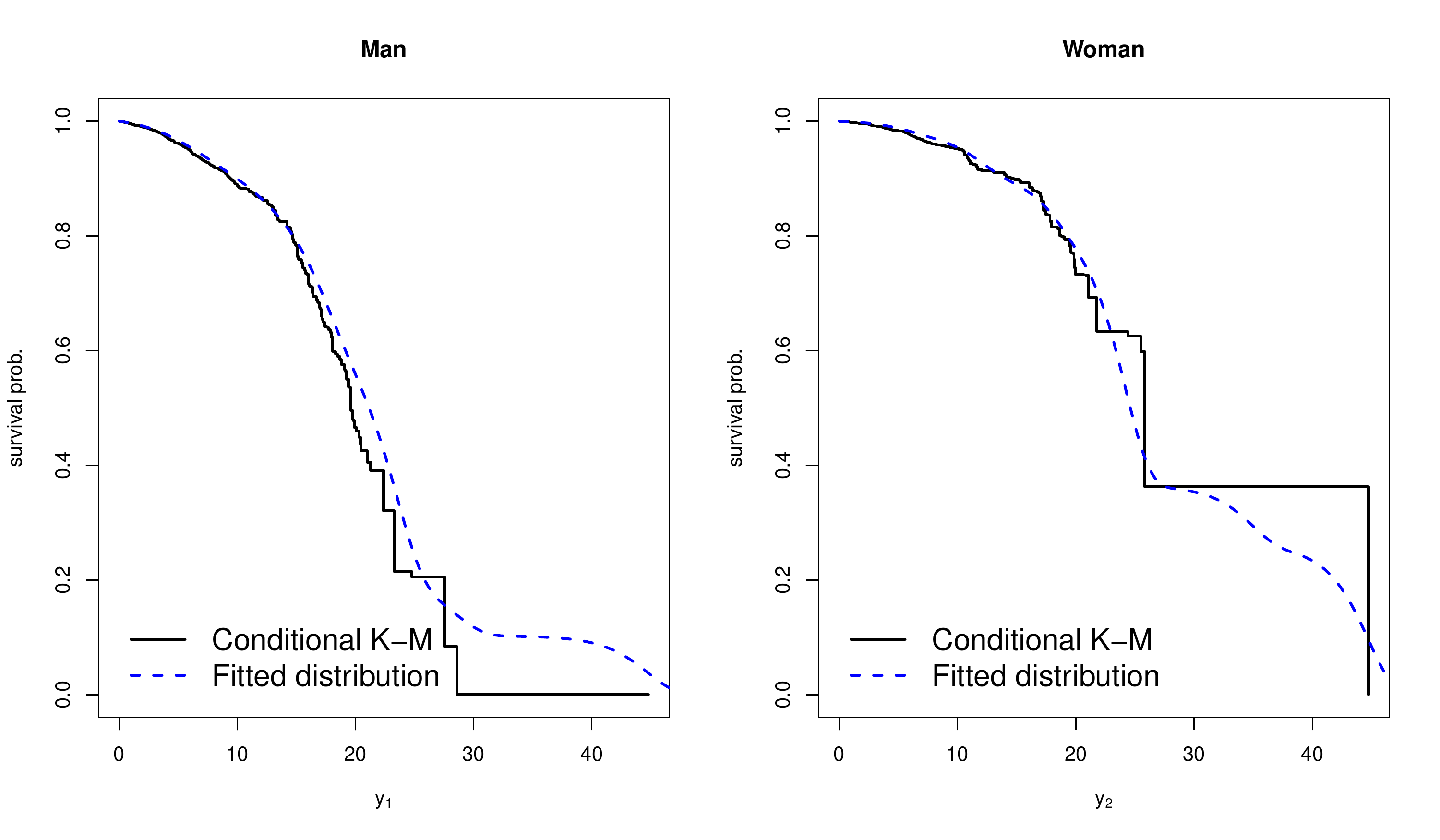}
        \caption{Couple 1}
    \end{subfigure}
    \vfill
        \begin{subfigure}[b]{0.9\textwidth}
        \includegraphics[width=1\textwidth]{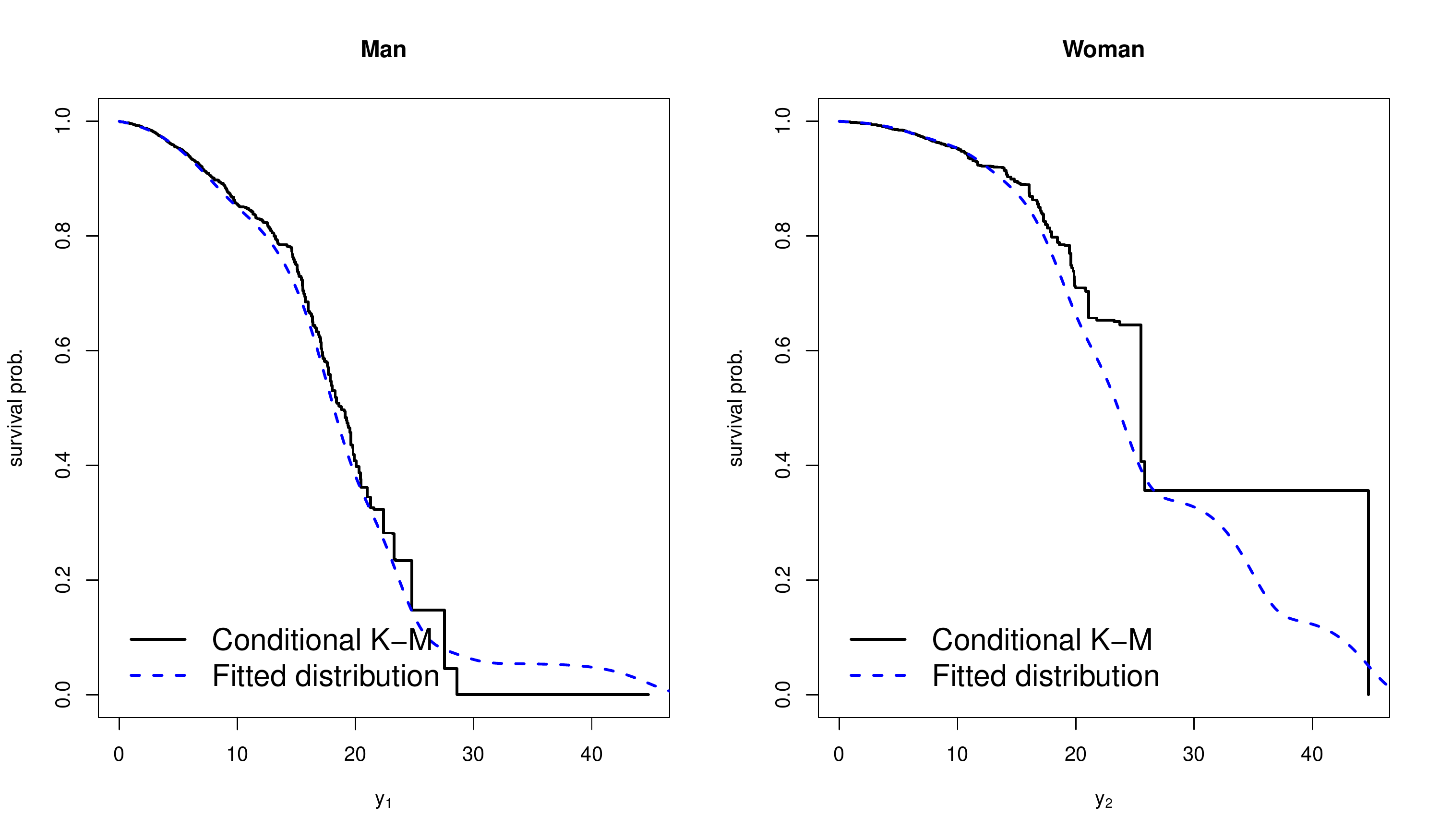}
        \caption{Couple 2}
    \end{subfigure}
    \caption{Conditional K-M estimators vs the fitted distribution, Couples 1 and 2.}
    \label{fig: beran comp c1 c2}
\end{figure}

\begin{figure}[ht]
    \centering
    \begin{subfigure}[b]{0.9\textwidth}
        \includegraphics[width=1\textwidth]{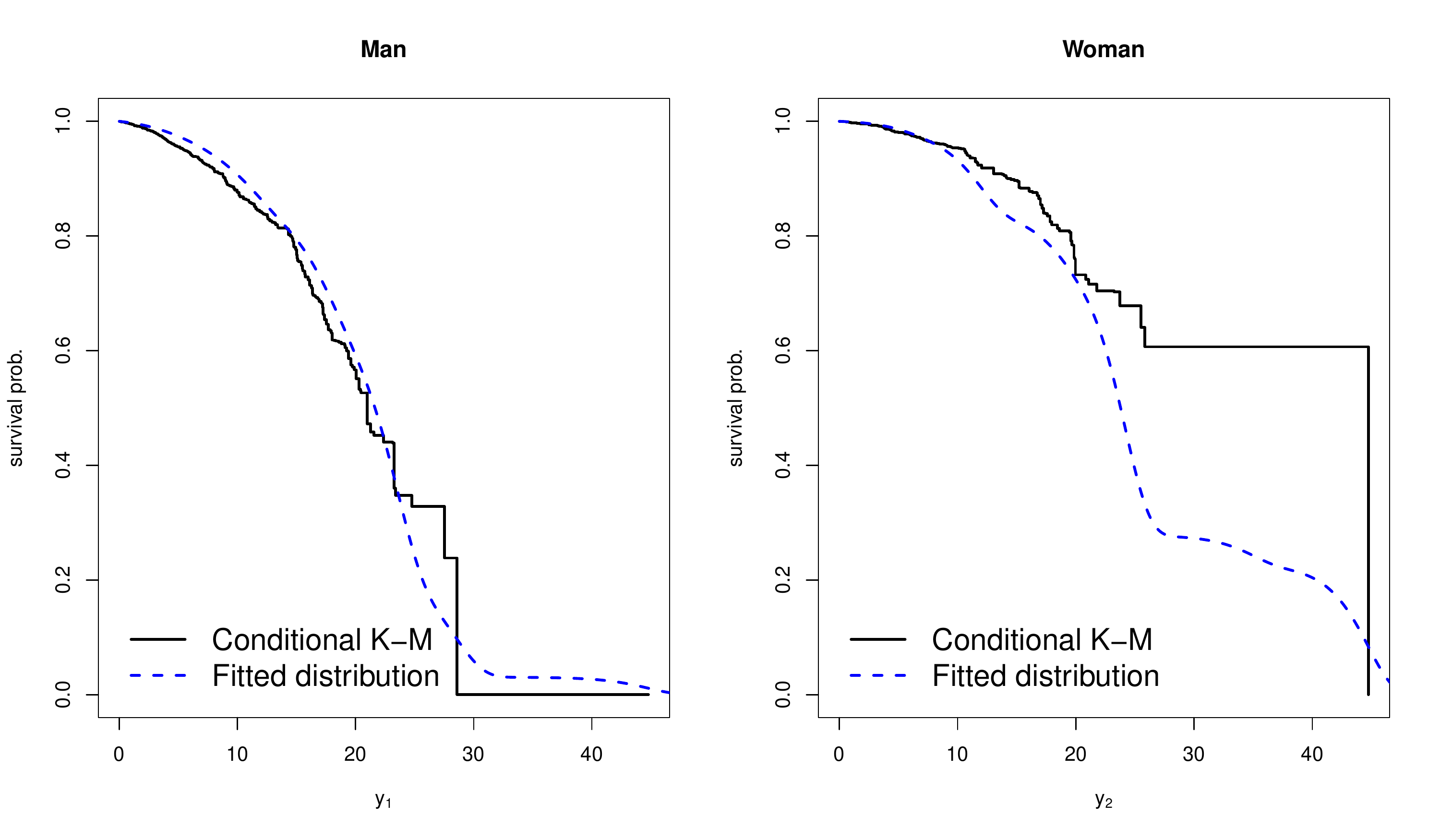}
        \caption{Couple 3}
    \end{subfigure}
    \vfill
        \begin{subfigure}[b]{0.9\textwidth}
        \includegraphics[width=1\textwidth]{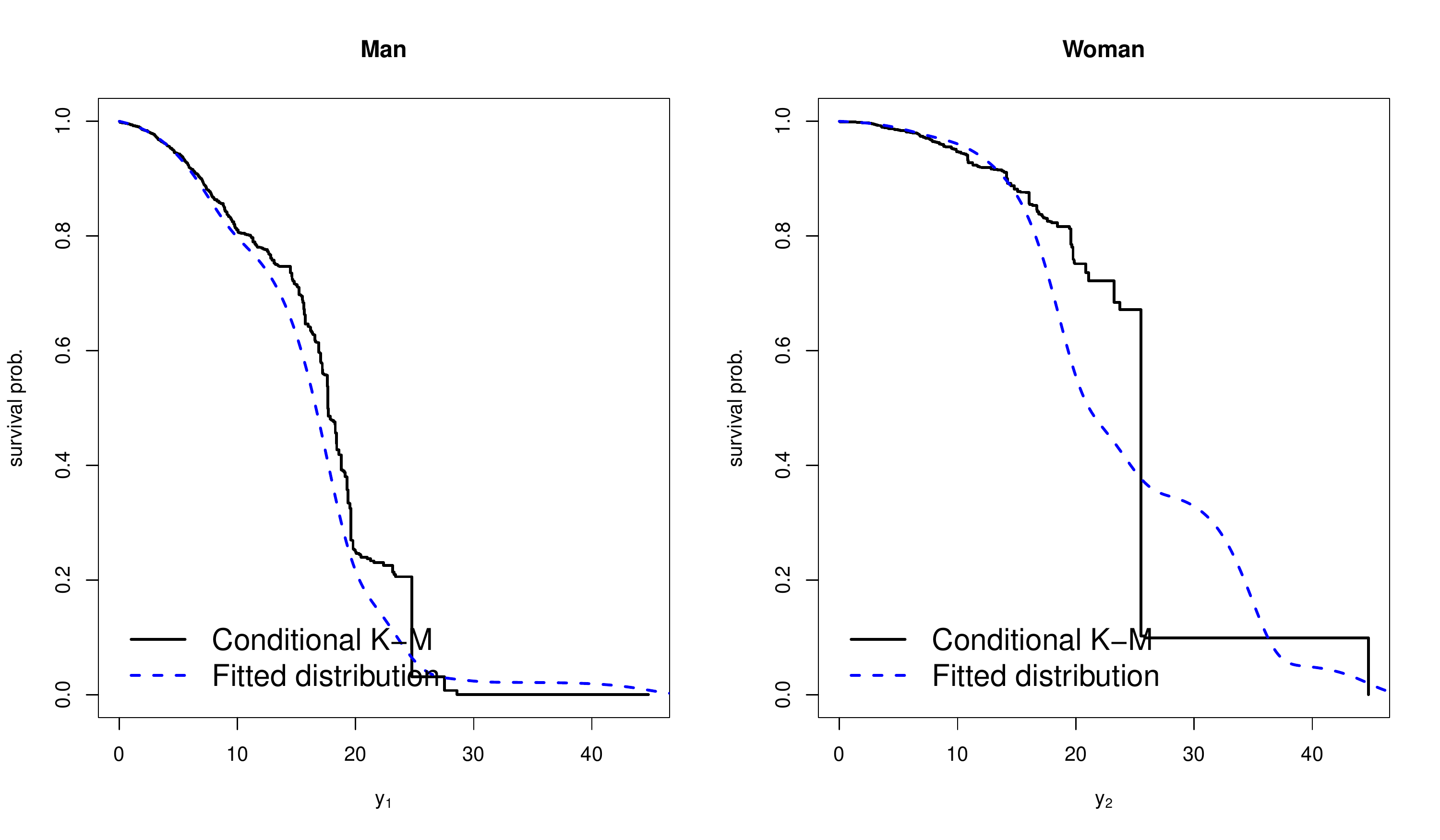}
        \caption{Couple 4}
    \end{subfigure}
    \caption{Conditional K-M estimators vs the fitted distribution, Couples 3 and 4.}
    \label{fig: beran comp c3 c4}
\end{figure}

To assess the precision of our estimated marginals $Y_i^c\sim\mbox{IPH}(\hat \bfp,\hat \bfT_i, \hat \beta_i)$, $i=1,2$ and $c=1,2,3,4$, we use the conditional Kaplan-Meier (K-M in the following) estimator, also known as Beran estimator. For a sample $X_1,X_2,\dots,X_n$ and covariate matrix $\mat{A}$, the conditional K-M estimator we use is 
\begin{align}\label{beran est}
    \begin{split}
         \hat \P(X\leq t\mid \mat{A}=\vect{a})&=\hat F_X(t\mid a)=1-\prod_{i:X_{i:n}\leq t}\left(1-\frac{\delta_{i:n} K\left(\frac{\vect{a}-\mat{A}_{i:n}}{b_n}\right)}{\sum_{j=i}^{n}K\left(\frac{\vect{a}-\mat{A}_{j:n}}{b_n}\right)} \right),
    \end{split}
\end{align}
where $X_{i:n}$ are order statistics of the sample, $\delta_{i:n}$ the respective right-censoring indicators, $K(\cdot)$ is a kernel function and $b_n$ is a band sequence. In our instance, the kernel function is a multivariate Gaussian density and $b_n=0.001$. For more details on the conditional K-M estimator we refer the reader to Dabrowska \cite{dabrowska1989}. Figures \ref{fig: beran comp c1 c2} and \ref{fig: beran comp c3 c4} compare the survival probabilities obtained by the conditional K-M estimators with the one of the fitted distributions. One sees that in all cases the fit is in fact quite satisfactory.  
%%%%%%%%%%%%%%%%%%%%%%%%%%%%%%%%%%%%%%%%%%%%%%%%%%%%%%%%%%%%%%%%%%%%%%%%%%%%%%%%%%%%%%%%%%%%%%%%%%%
%%%%%%%%%%%%%%%%%%%%%%%%%%%%%%%%%%%%%%%%%%%%%%%%%%%%%%%%%%%%%%%%%%%%%%%%%%%%%%%%%%%%%%%%%%%%%%%%%%
\subsection{Joint behaviour}
\begin{figure}[ht]
    \centering
    \begin{subfigure}[b]{0.495\textwidth}
        \includegraphics[width=1\textwidth]{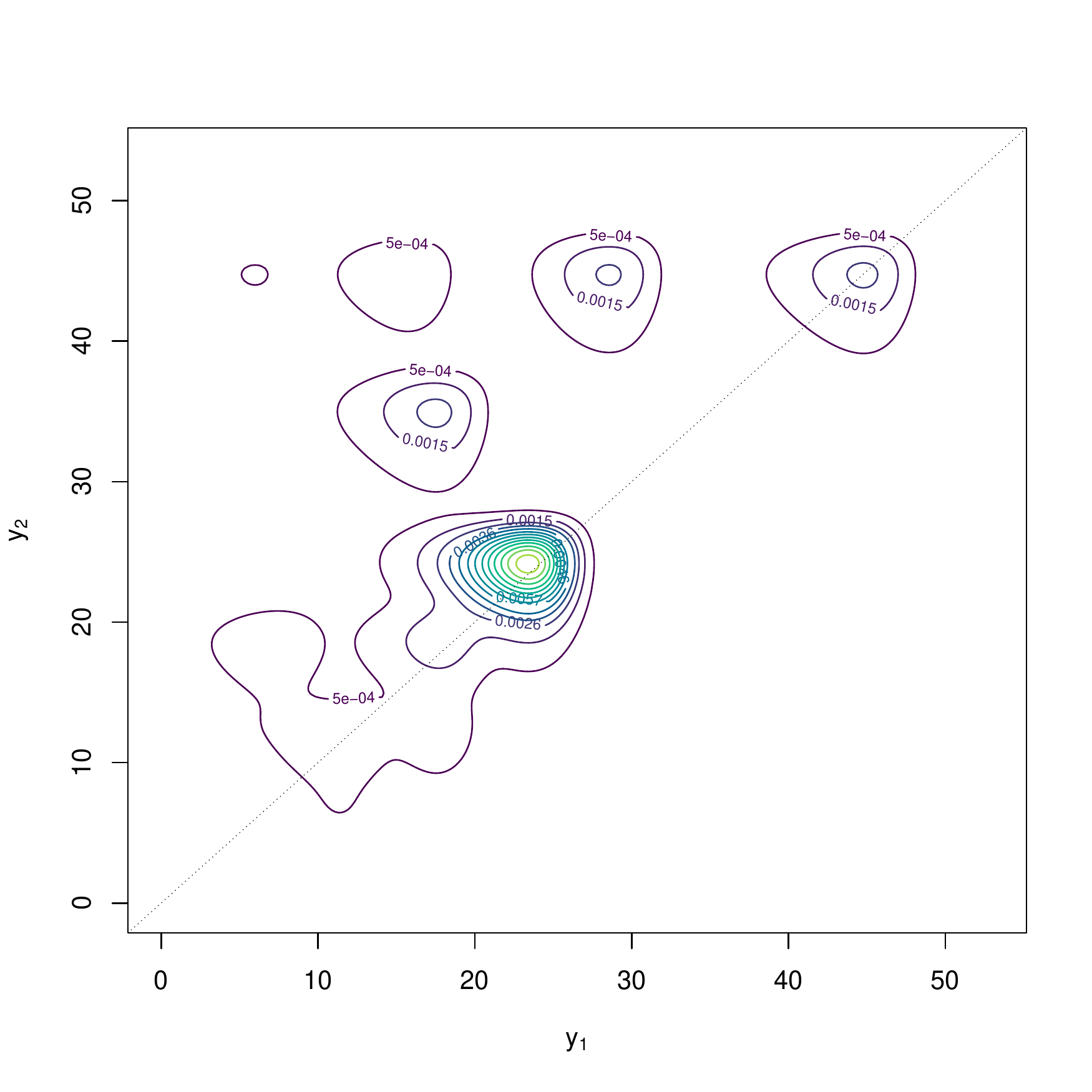}
        \caption{Couple 1}
    \end{subfigure}
    \hfill
        \begin{subfigure}[b]{0.495\textwidth}
        \includegraphics[width=1\textwidth]{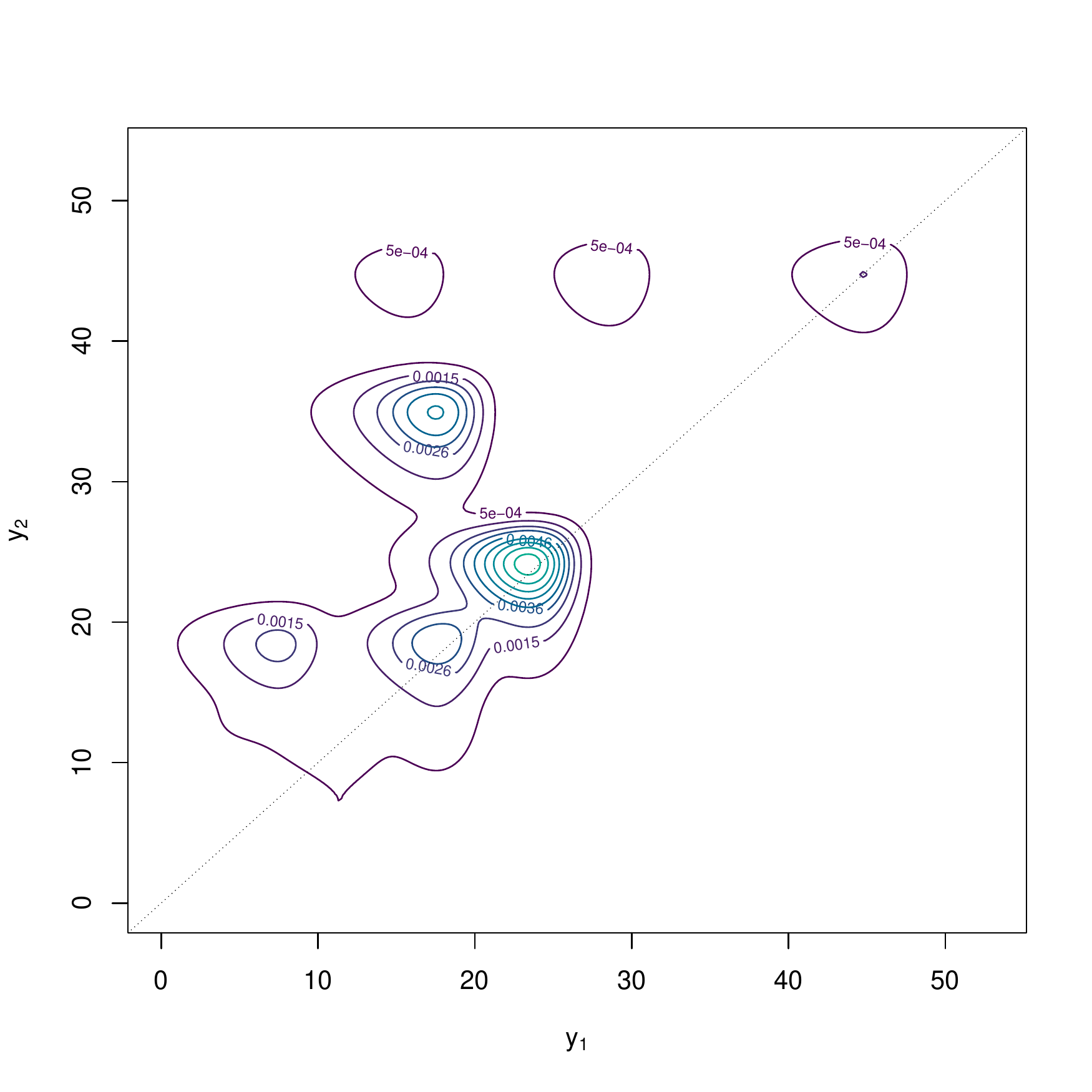}
        \caption{Couple 2}
    \end{subfigure}
    \vfill
        \begin{subfigure}[b]{0.495\textwidth}
        \includegraphics[width=1\textwidth]{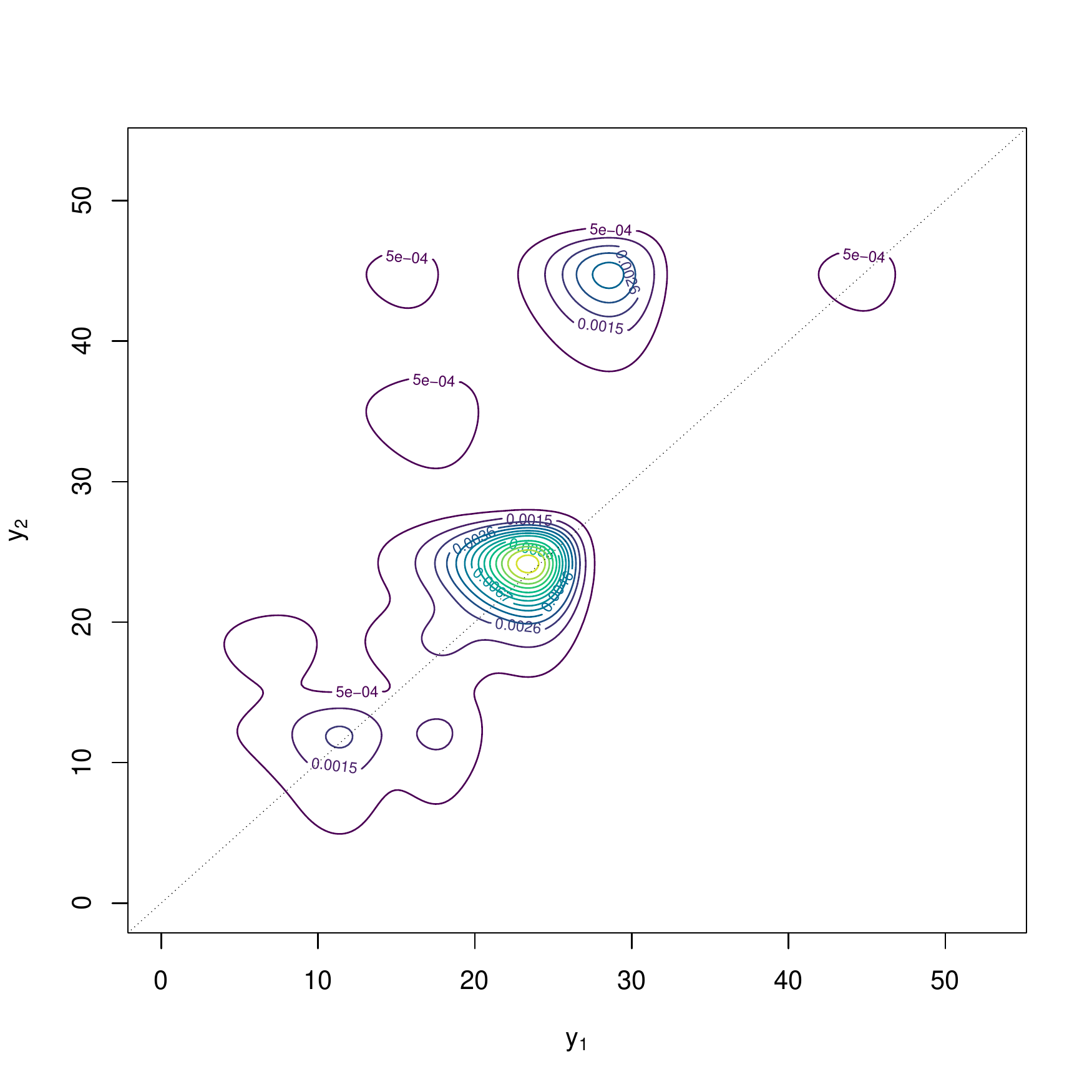}
        \caption{Couple 3}
    \end{subfigure}
    \hfill
        \begin{subfigure}[b]{0.495\textwidth}
        \includegraphics[width=1\textwidth]{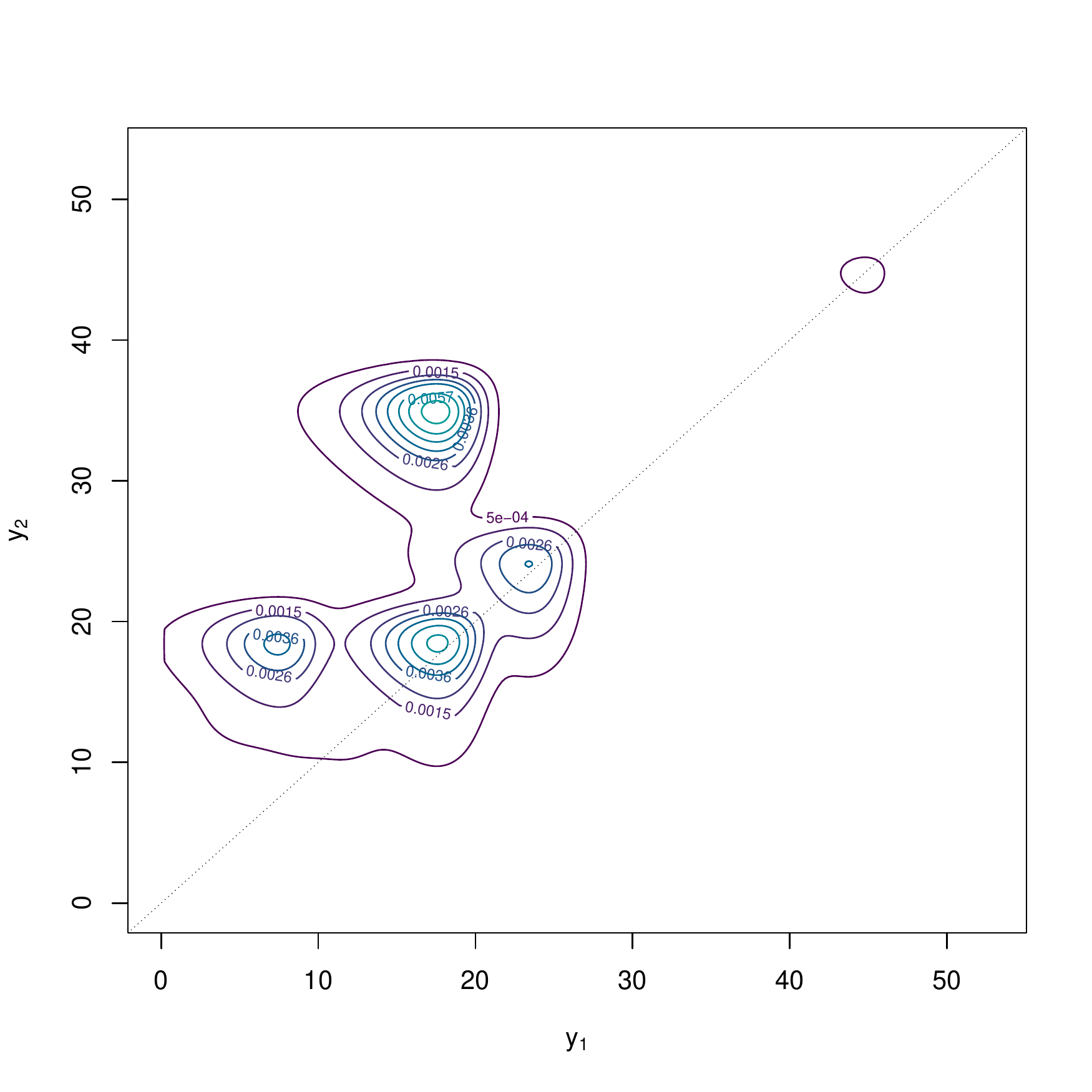}
        \caption{Couple 4}
    \end{subfigure}
    \caption{Contour plots of the bivariate lifetime densities for the four couples}
    \label{fig: joint pdf comp}
\end{figure}

Let us now consider the resulting bivariate densities. 
In Figure \ref{fig: joint pdf comp} we find the bivariate densities for the four specified couples.
The joint density of Couple 1 shows that large differences in their survival times are quite unlikely (the vast majority of the joint mass being located close the identity line). Also, the most likely survival times are close to $23$ years. There is also a considerable probability mass above the identity line, where the woman survives longer than the man. For instance, we have $\Bar{F}_{\mat{Y}^1}(12,30)=32\%$ while $\Bar{F}_{\mat{Y}^1}(30,12)=11.79\%$.\\ 
For Couple 2 the situation is different (top right panel of Figure \ref{fig: joint pdf comp}): with the man being already 68 years and the woman being 63 years old, the remaining lifetimes are shorter, with the major mode of the joint distribution being located near $(y_1,y_2)=(23,24)$, the second largest close to $(y_1,y_2)=(17,35)$ and the next ones in the neighbourhood of $(y_1,y_2)=(17,19)$ and $(y_1,y_2)=(9,19)$, respectively. There is now a much higher probability for the husband to die sooner.\\
The joint density for Couple 3 resembles the one of Couple 1, with survival times beyond $40$ years being more unlikely. 
\textcolor{black}{Despite the man being the same age as the man in Couple 1, there is a larger probability for the couple to have survival times close to $(y_1,y_2)=(29,45)$ than for Couple 1.}\\
Finally, the joint density of Couple 4 is close to the one of Couple 2. The spike around $(y_1,y_2)=(17,35)$ is more pronounced than the analogue of  Couple 2, while the spike close to $(y_1,y_2)=(23,24)$ is less important than its counterpart in Couple 2.\\
One may be tempted to conclude from these four distributions that for the same age, having a younger partner leads to longer survival times, which would signal that the bereavement effect is weaker when spouses have larger differences in age at issue of a policy. 
\subsection{Dependence measures}
\begin{figure}[ht]
    \centering
    \begin{subfigure}[b]{0.495\textwidth}
        \includegraphics[width=1\textwidth]{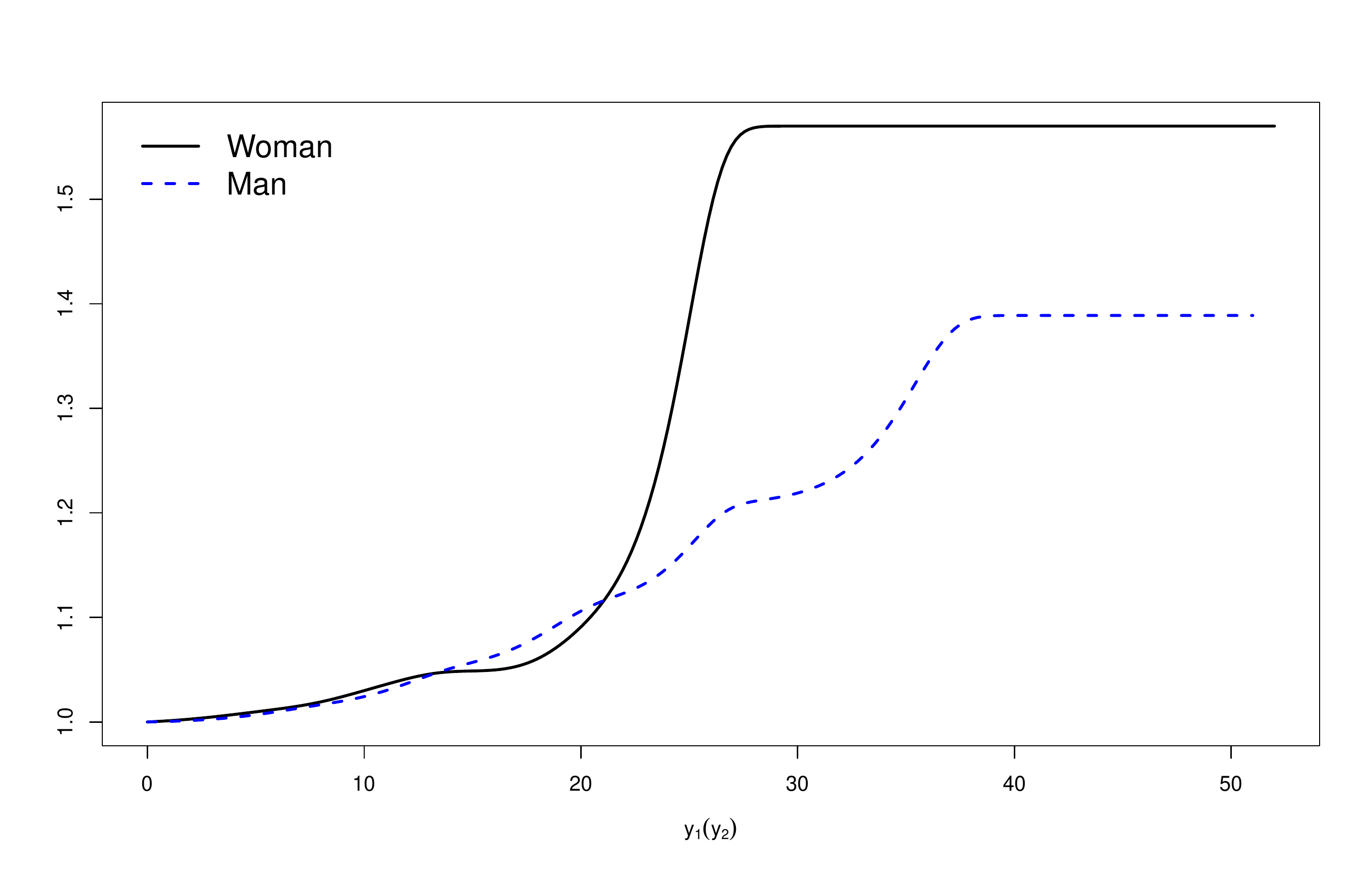}
        \caption{Couple 1}
    \end{subfigure}
    \hfill
        \begin{subfigure}[b]{0.495\textwidth}
        \includegraphics[width=1\textwidth]{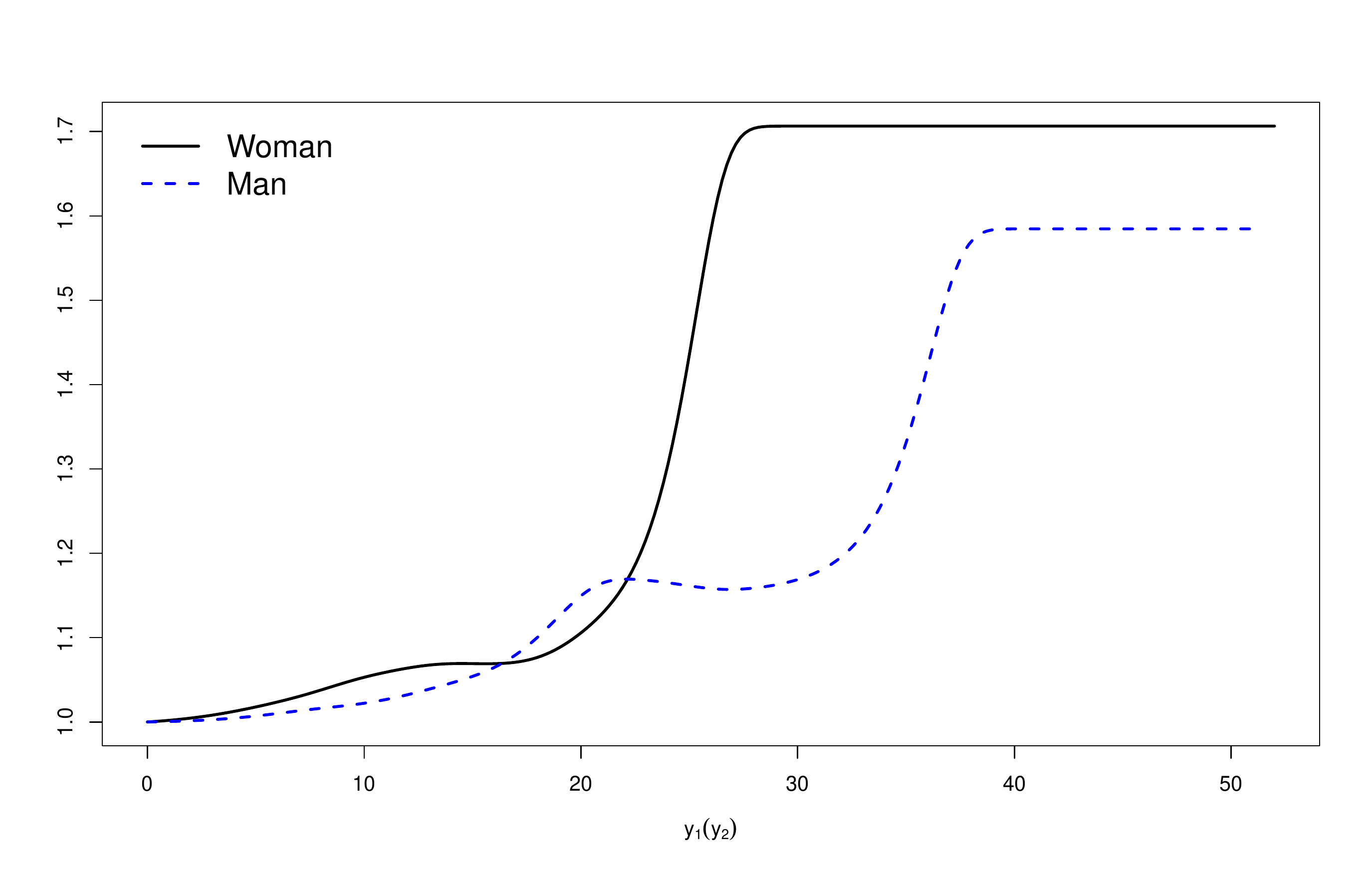}
        \caption{Couple 2}
    \end{subfigure}
    \vfill
        \begin{subfigure}[b]{0.495\textwidth}
        \includegraphics[width=1\textwidth]{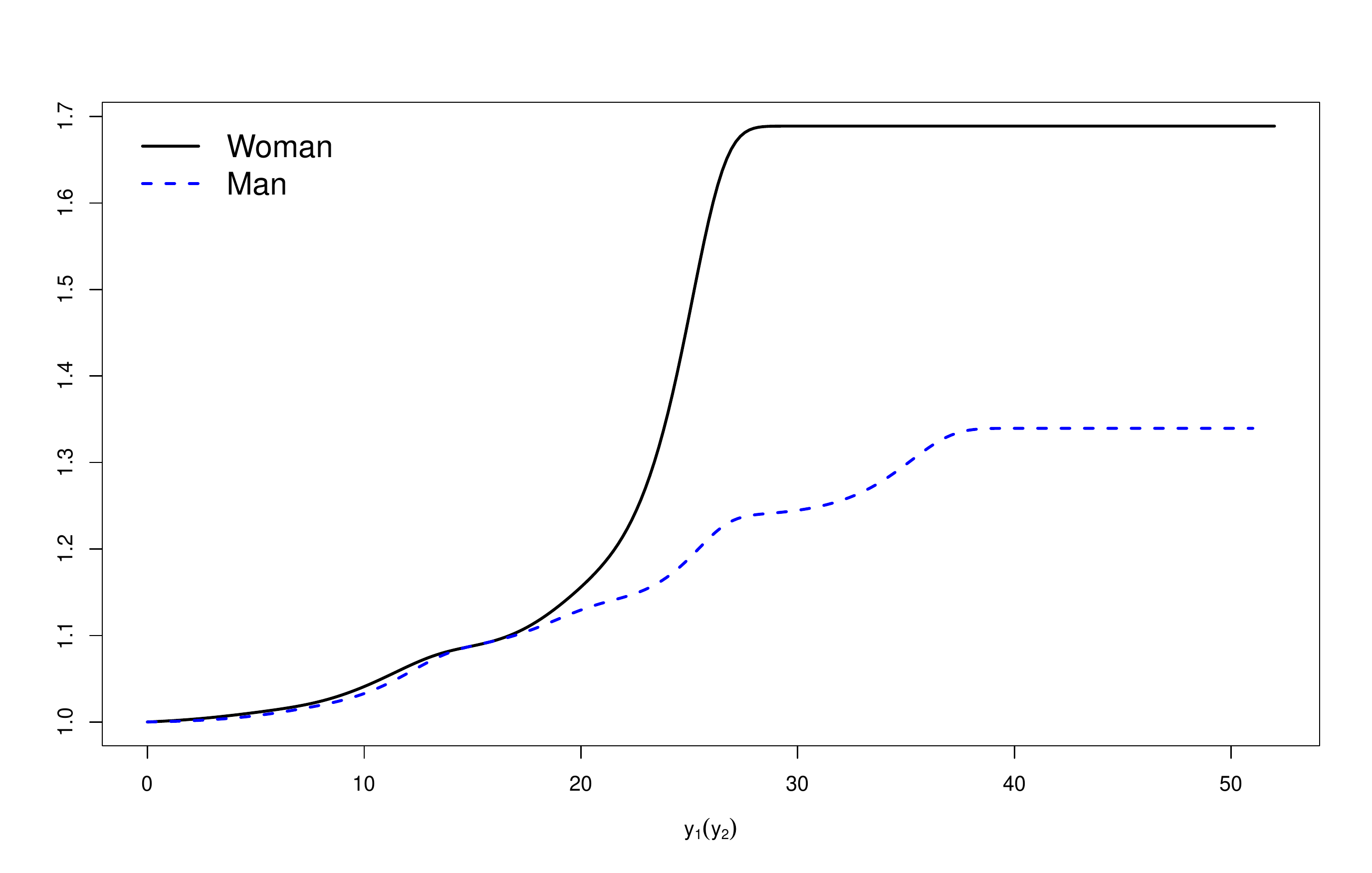}
        \caption{Couple 3}
    \end{subfigure}
    \hfill
        \begin{subfigure}[b]{0.495\textwidth}
        \includegraphics[width=1\textwidth]{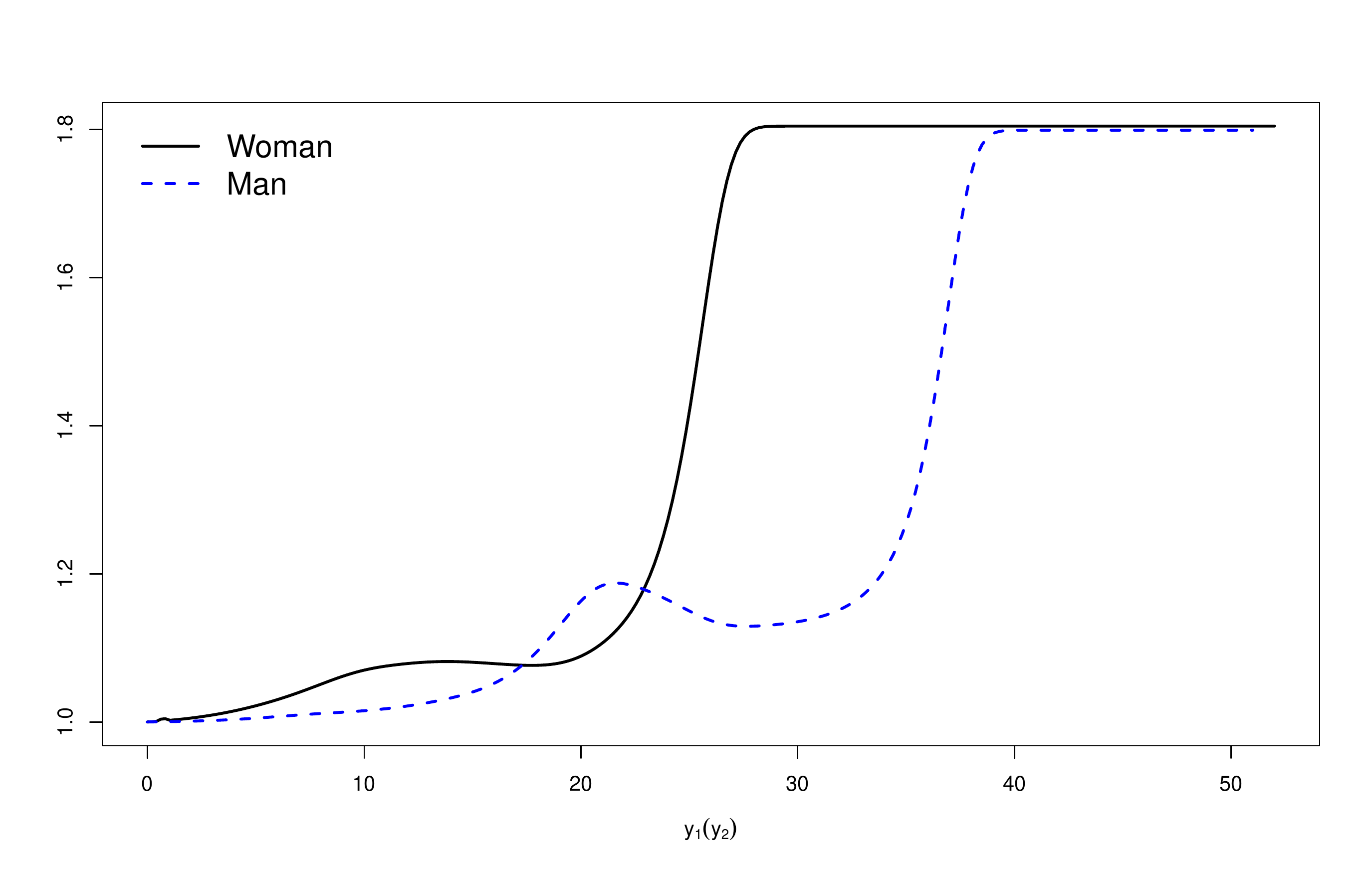}
        \caption{Couple 4}
    \end{subfigure}
    \caption{$\Psi_2^1(0,y)$ and $\Psi_2^2(x,0)$ for the four couples.}
    \label{fig: Phi2 comp}
\end{figure}

\begin{figure}[ht]
    \centering
    \begin{subfigure}[b]{0.495\textwidth}
        \includegraphics[width=1\textwidth]{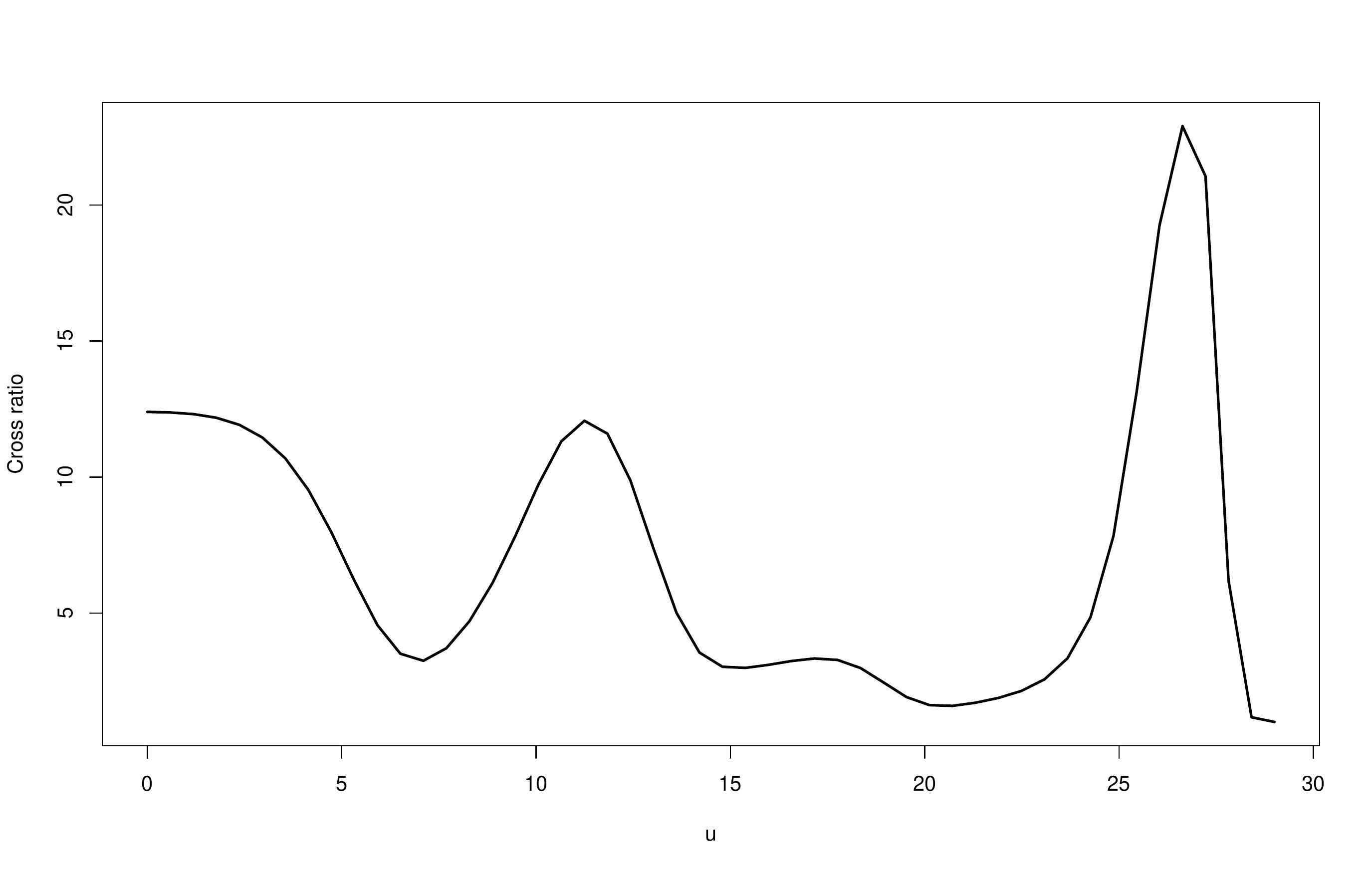}
        \caption{Couple 1}
    \end{subfigure}
    \hfill
        \begin{subfigure}[b]{0.495\textwidth}
        \includegraphics[width=1\textwidth]{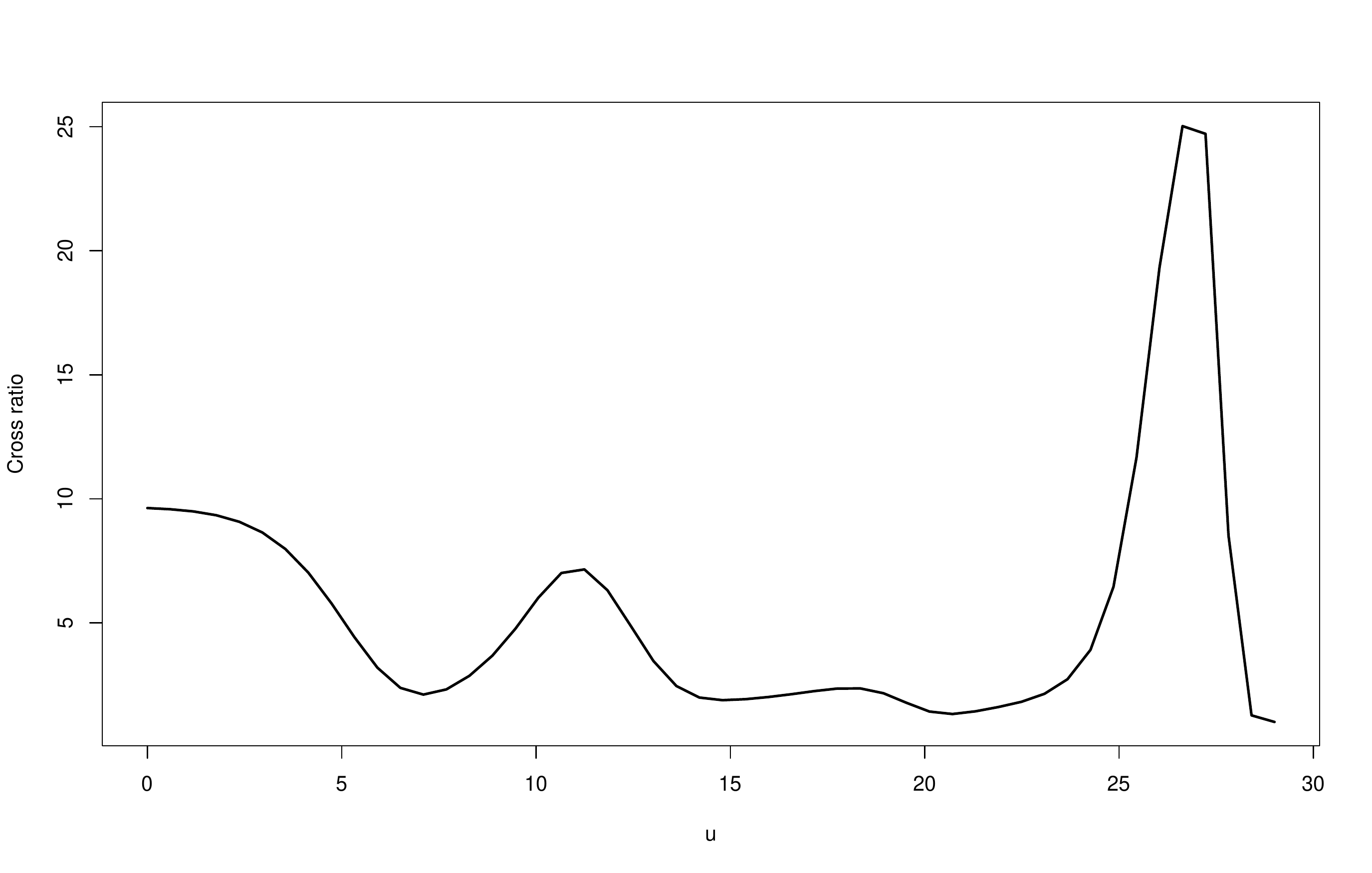}
        \caption{Couple 2}
    \end{subfigure}
    \vfill
        \begin{subfigure}[b]{0.495\textwidth}
        \includegraphics[width=1\textwidth]{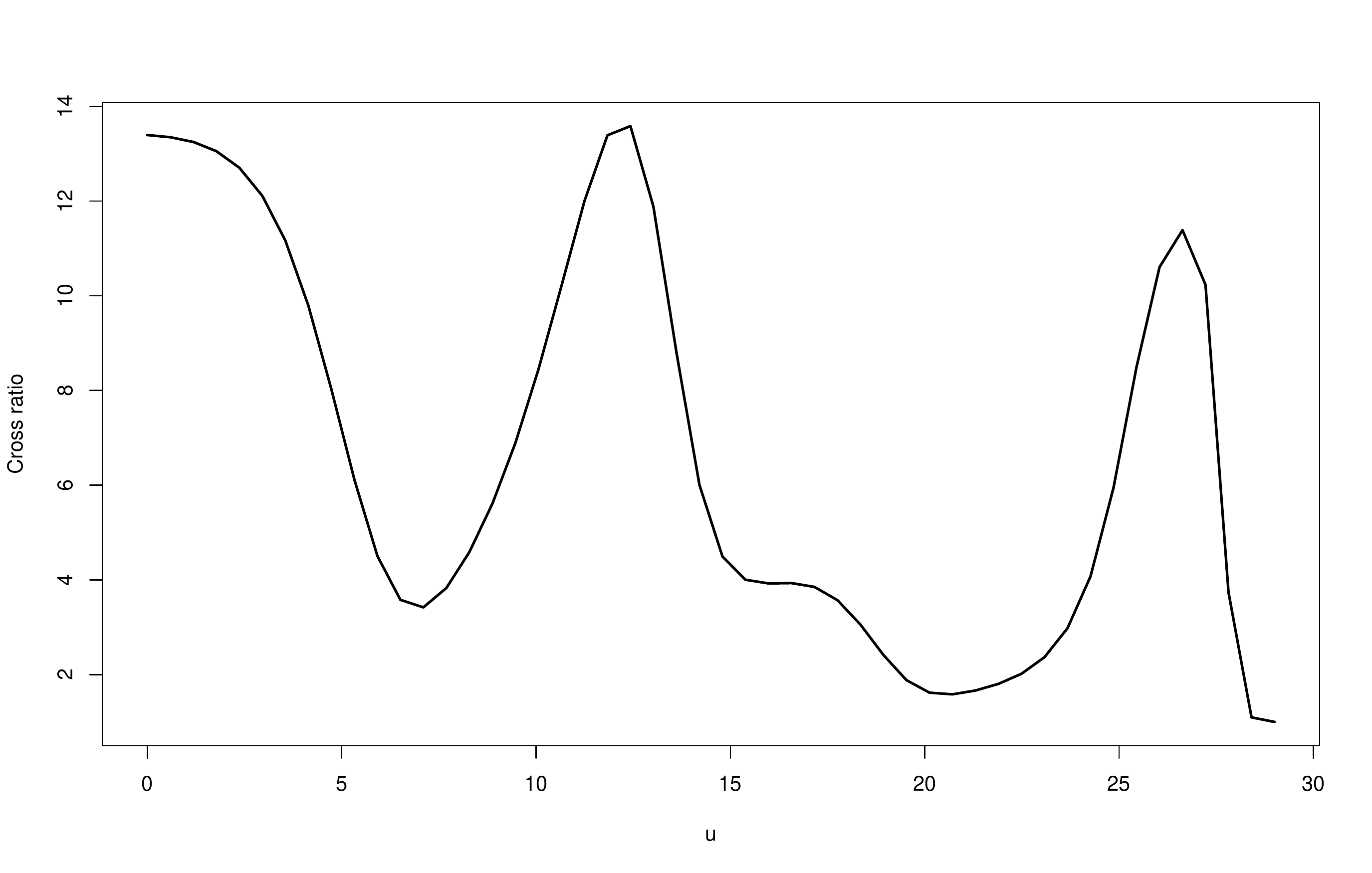}
        \caption{Couple 3}
    \end{subfigure}
    \hfill
        \begin{subfigure}[b]{0.495\textwidth}
        \includegraphics[width=1\textwidth]{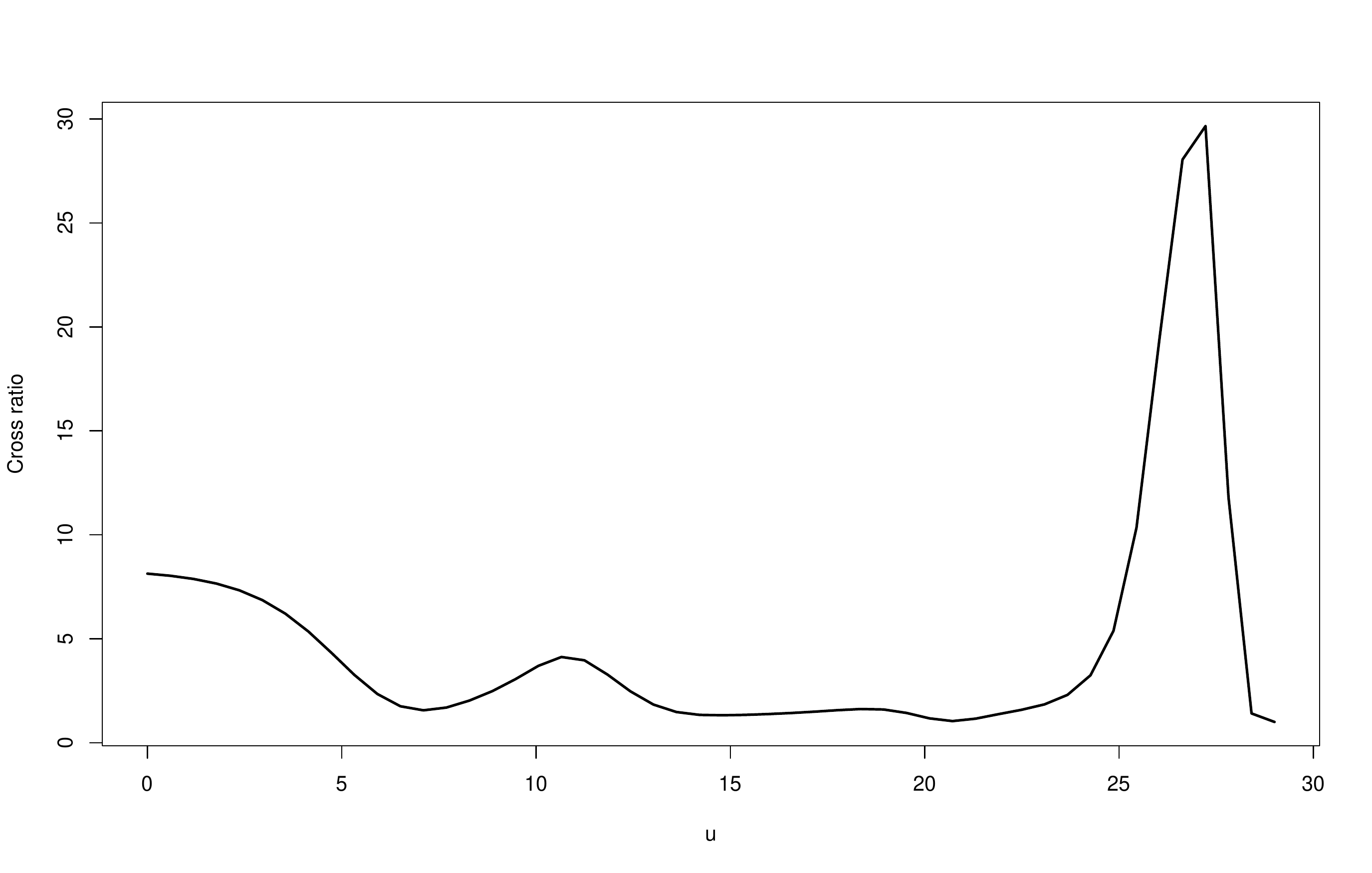}
        \caption{Couple 4}
    \end{subfigure}
    \caption{$CR(u,u)$ for the four couples.}
    \label{fig: CR comp}
\end{figure}

Let us now explore some dependence measures for these four exemplary couples. In addition to Kendall's tau and Spearman's rho, we also consider three measures of time-dependent association, which were analysed in Luciano et al.\ \cite{luciano2008}.\\
The mIPH distributions we consider all have the same copula as the corresponding mPH distributions with equal representation, since matrix-Gompertz distributions are monotone increasing transformations of PH distributions. At the same time, for the mPH class we have explicit expressions for Kendall's tau and Spearman's rho. The pairwise Kendall's tau of marginals $X_k$ and $X_l$ is 
\[\tau_{X_k,X_l}=4\sum_{i=1}^p\sum_{j=1}^p\pi_i\pi_j   (\bfe_i^\mathsf{T}\otimes \bfe_j^\mathsf{T}) [-\bfT_k  \oplus \bfT_k ]^{-1} (\bfe\otimes\bft_k)
   (\bfe_i^\mathsf{T}\otimes \bfe_j^\mathsf{T})[-\bfT_l  \oplus \bfT_l ]^{-1}(\bfe\otimes\bft_l)-1,
\]
while Spearman's rank correlation is given as
\[\rho^S_{X_k,X_l}=12\sum_{j=1}^p\pi_j
\left(1-(\bfp\otimes \bfe_j^{\mathsf{T}})[-\bfT_k \oplus \bfT_k]^{-1}(\bfe \otimes \bft_k)\right)
 \left(1-(\bfp\otimes \bfe_j^{\mathsf{T}})[-\bfT_l \oplus \bft_l]^{-1}(\bfe\otimes \bfT_l)\right)-3,
\]
cf.\ \cite{bladt2022tractable}. Calculating these quantities for the four specified couples, one obtains
\begin{align*}
    \tau_{Y_1^1,Y_2^1}&=0.3104, \quad \tau_{Y_1^2,Y_2^2}=0.2562, \quad \tau_{Y_1^3,Y_2^3}=0.4367, \quad \tau_{Y_1^4,Y_2^4}=0.2139, \\
    \rho_{Y_1^1,Y_2^1}^S&=0.4526, \quad \rho_{Y_1^2,Y_2^2}^S=0.3938, \quad \rho_{Y_1^3,Y_2^3}^S=0.6144, \quad \rho_{Y_1^4,Y_2^4}^S=0.3381.
\end{align*}
All couples manifest positive concordance, which on top of Figure \ref{fig: joint pdf comp} is additional evidence that the lifetimes of individuals in a couple are correlated. We find the strongest concordance in couples where women are older than their husbands. Concretely, Couple 3 has the highest Kendall's tau and Spearman's rho values, followed by Couple 1. Couple 2  and Couple 4 have lower, and similar, corresponding values.\\
\begin{figure}[ht]
    \centering
    \begin{subfigure}[b]{0.495\textwidth}
        \includegraphics[width=1\textwidth]{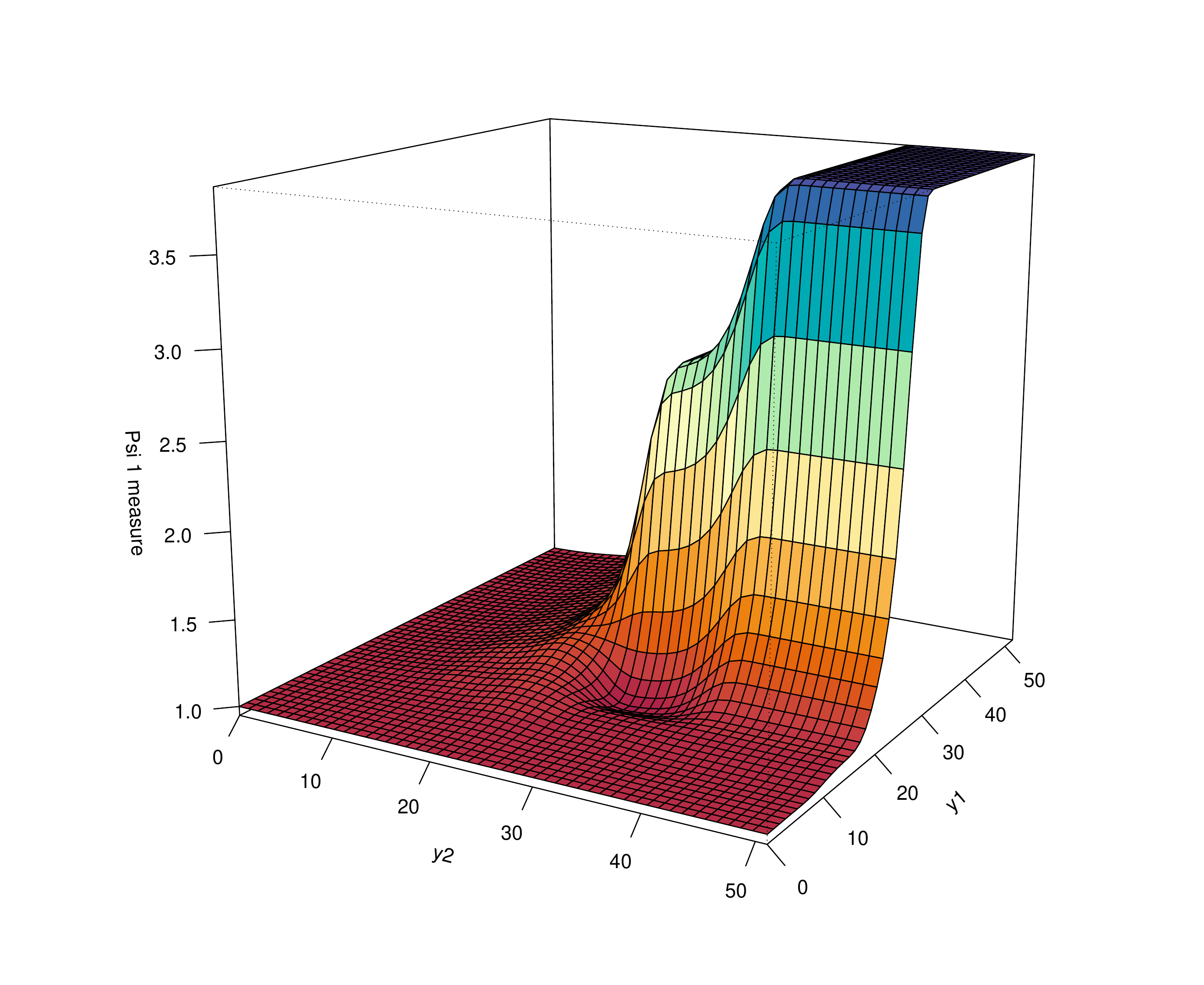}
        \caption{Couple 1}
    \end{subfigure}
    \hfill
        \begin{subfigure}[b]{0.495\textwidth}
        \includegraphics[width=1\textwidth]{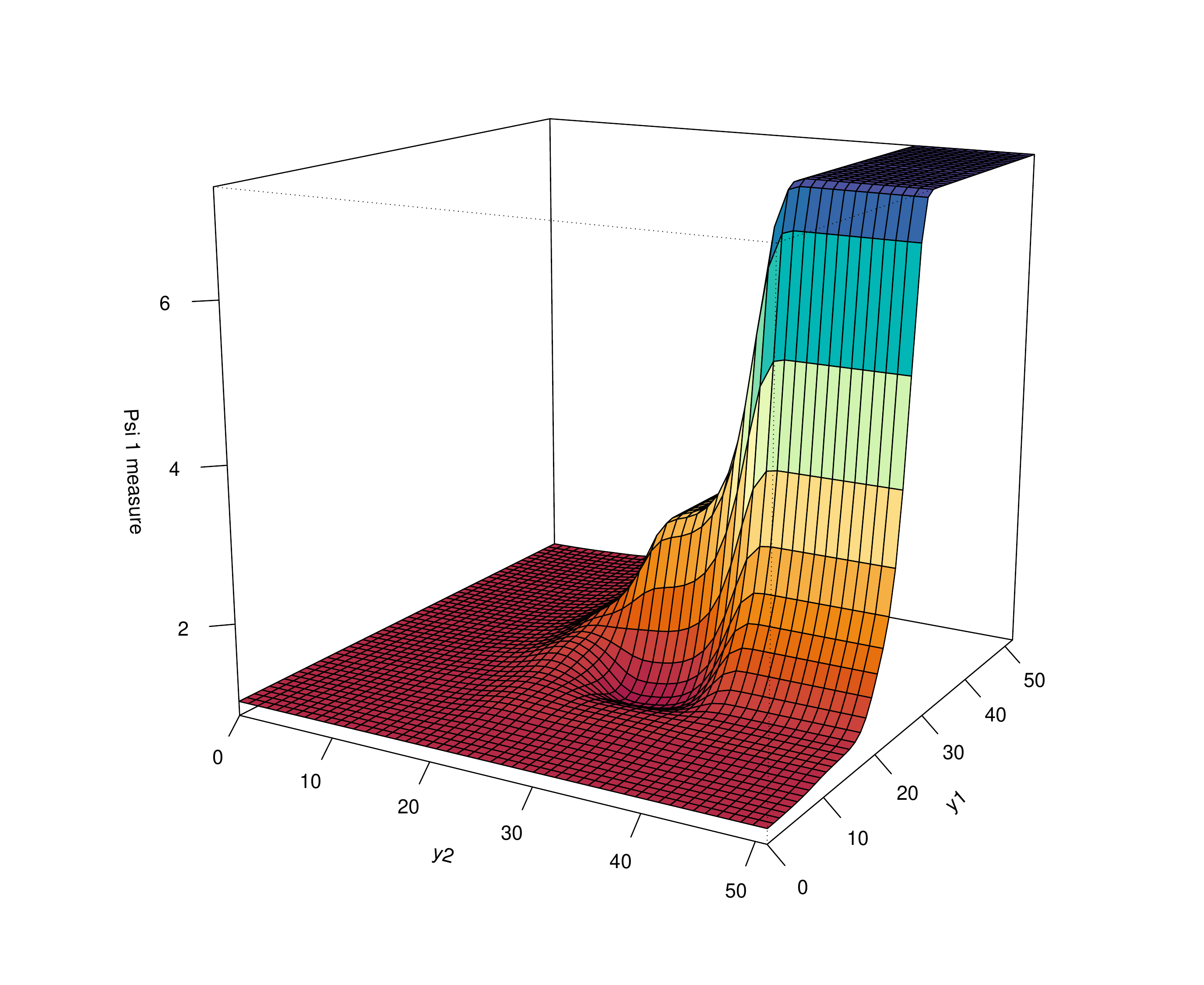}
        \caption{Couple 2}
    \end{subfigure}
    \vfill
        \begin{subfigure}[b]{0.495\textwidth}
        \includegraphics[width=1\textwidth]{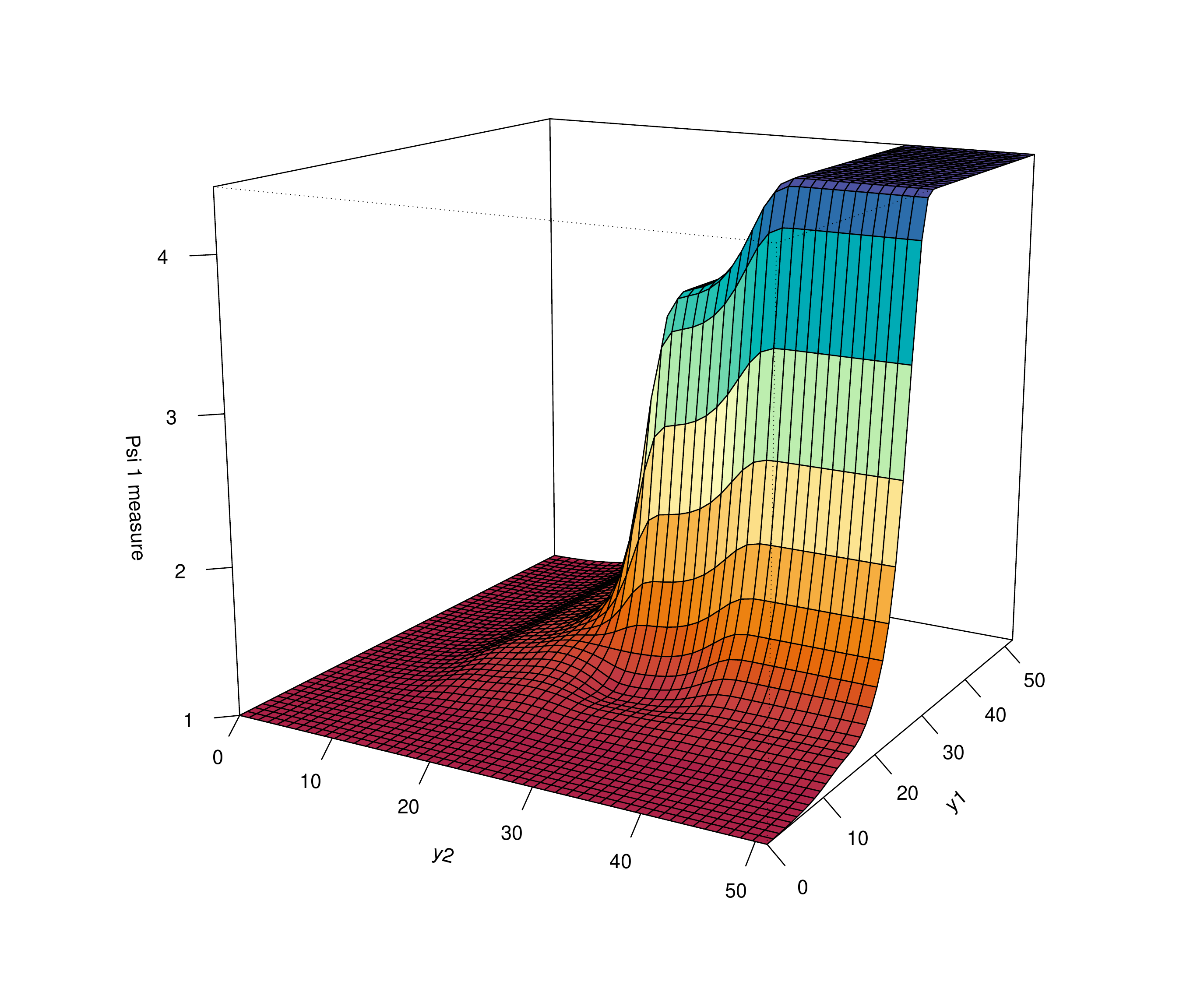}
        \caption{Couple 3}
    \end{subfigure}
    \hfill
        \begin{subfigure}[b]{0.495\textwidth}
        \includegraphics[width=1\textwidth]{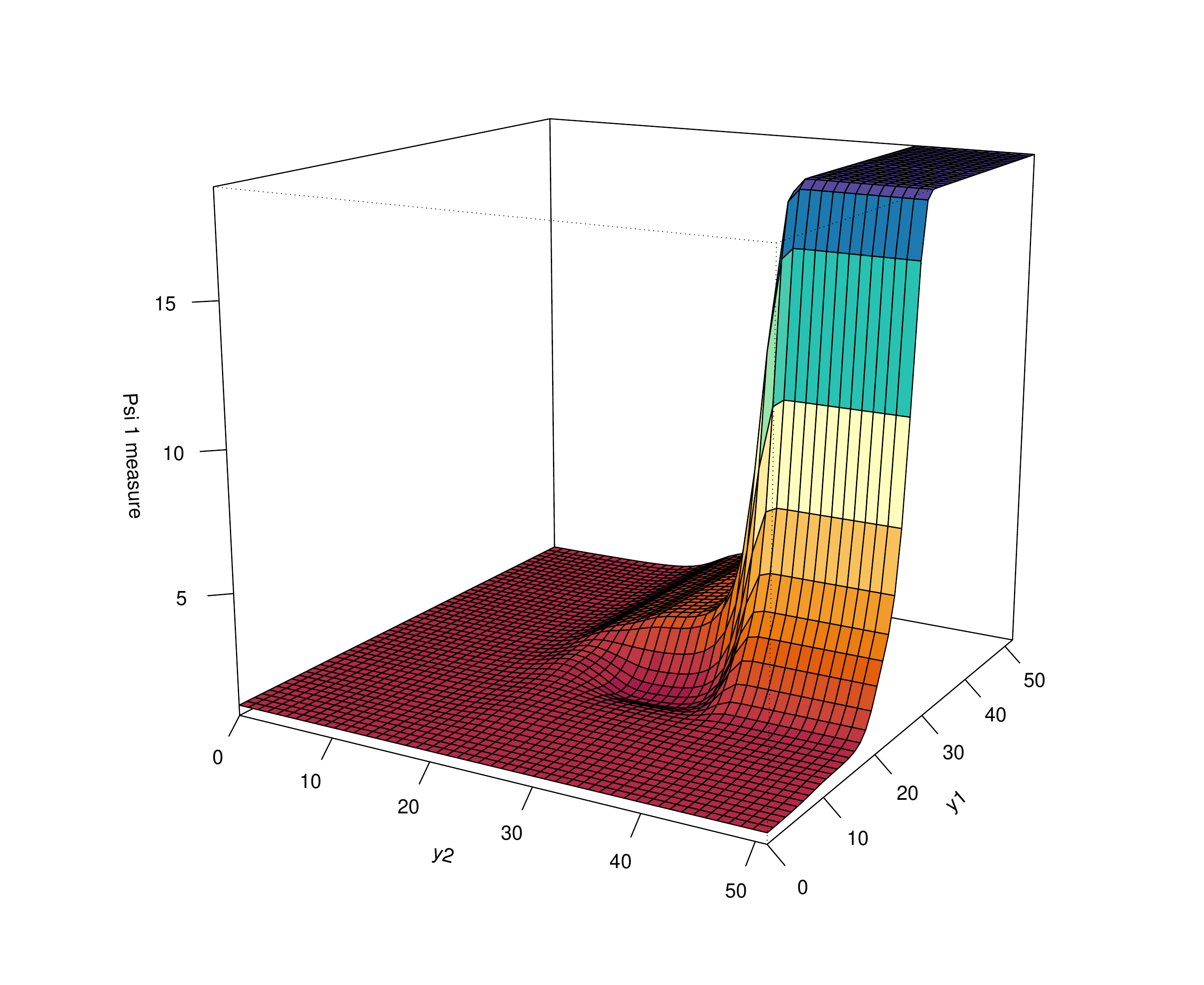}
        \caption{Couple 4}
    \end{subfigure}
    \caption{$\Psi_1(y_1,y_2)$ for the four couples.}
    \label{fig: Phi1 comp}
\end{figure}
As a first measure of time-dependent association, we consider in Figure \ref{fig: Phi1 comp} $$\Psi_1(y_1,y_2)=S(y_1,y_2)/\left(S_{Y_1}(y_1)S_{Y_2}(y_2)\right).$$
Most of the resulting values are greater than $1$, indicating positive dependence. However, for values of $y_2$ close to $30$, Couples 1,2 and 3 exhibit $\Psi_1(y_1,y_2)<1$. This suggests negative dependence for remaining lifetimes when women survive at least 30 years and men at least 20 years. One can see that, roughly, after values $y_1=29$ and $y_2=39$, the ratios $\Psi_1(y_1,y_2)$ remain constant. This happens since for very large survival times $y_1,y_2$, marginal survival probabilities change very little, and this change is absorbed by $S(y_1,y_2)$.\\
Next, we depict in Figure \ref{fig: Phi2 comp} the measures
\begin{align*}
    \Psi_2^1(0,y_2)&=\E\left(Y_1 \mid Y_2 \ge y_2\right)/\,\E\left(Y_1 \right),\\
     \Psi_2^2(y_1,0)&=\E\left(Y_2 \mid Y_1 \ge y_1 \right)/\,\E\left(Y_2 \right)
\end{align*}
for all four couples, 
which give the relative change of conditional expectations of spouses' excess lifetimes, given that the partner survives at least $y_2(y_1)$ years. We see that the latter increase throughout with $y_2(y_1)$. In general, the ratios $\Psi_2^i(\cdot,\cdot)$, $i=1,2$, are close in value when $y_1,y_2\le 20$, meaning that the survival of a spouse has a similar effect on the remaining lifetime of the partner for the first 20 years. After that, the relative lifetime improvement increases much faster for women, i.e.\ their expected lifetime improvement is then more sensitive to the survival of the partner than vice versa. Like in Figure \ref{fig: Phi1 comp}, both ratios remain constant after spouses' survival times of $y_1=29$ for women and $y_2=39$ for men.\\
The last measure of time-dependent association we consider here is the cross-ratio $$CR(y_1,y_2)=S(y_1,y_2)\frac{\frac{d^2}{dy_1 dy_2}S(y_1,y_2)}{\frac{d}{dy_1}S_{Y_1}(y_1)\frac{d}{dy_2}S_{Y_2}(y_2)}$$
originally introduced by Clayton \cite{clayton1978model}, which gives the relative increase of the force of mortality of an individual immediately after death of the partner. The quantity relevant in our model is $CR(u,u)$, and Figure \ref{fig: CR comp} depicts the resulting figures for our model. 
In Luciano et al.\ \cite{luciano2008}, the resulting curves were monotone increasing in $u$, as a result of the imposed copula assumption on the joint lifetimes. In contrast, in the present setup these curves are not monotone increasing in $u$. One may interpret that in  the present approach the a priori dependence assumptions are less specified, and the data have a stronger impact on the resulting shape of $CR(u,u)$ than in a specified copula model, where the data merely influence the value of the dependence parameter. One can see from Figure \ref{fig: CR comp} that the cross-ratios exceed $1$ for $u\le 29$, so the survivor's force of mortality is  increased immediately after the death of the spouse, showing a bereavement effect (or broken-heart syndrom), but not a monotone one (with the magnitude  varying across age combinations of the couple). 
This bereavement effect somewhat seems to disappear for survival times beyond 29 years, and for this reason we do not plot values $u>29$, but in that range survival probabilities are very low anyway, and there are very few data points in that range to draw strong conclusions. 

Eventually, like in many other situations, it may depend on the number of available data points whether one prefers to have a flexible dependence structure in the fitting or a pre-specified copula family with possibly attractive stylized features, especially for extrapolated conclusions in regions with few data points. In this discussion, one may still appreciate the immediate causal interpretation of the mIPH model in terms of a common ageing mechanism. 

%Using definitions of short-term and long term dependence found in Spreeuw \cite{spreeuw2006}, which were used on the same data-set in Spreuw \& Owadally \cite{spreeuw_owadally_2013}, we shall determine which kind of dependence is present in our model. Given Property \ref{cond_dist}, the conditional distribution of $Y_i$, given that $Y_j$ survives exactly $t$ years, is again IPH with new initial distribution vector depending on $t$, for $i,j=1,2$ and $i\neq j$.

% This is not surprising since joint densities of $\mat{Y}^1$ and $\mat{Y}^3$ have their mass closer to the identity line. On the contrary, joint distributions of Couple 2 and 4 allow for more divergent survival times.
\subsection{Life expectancies}
\begin{figure}[ht]
    \centering
    \begin{subfigure}[b]{0.495\textwidth}
        \includegraphics[width=1\textwidth]{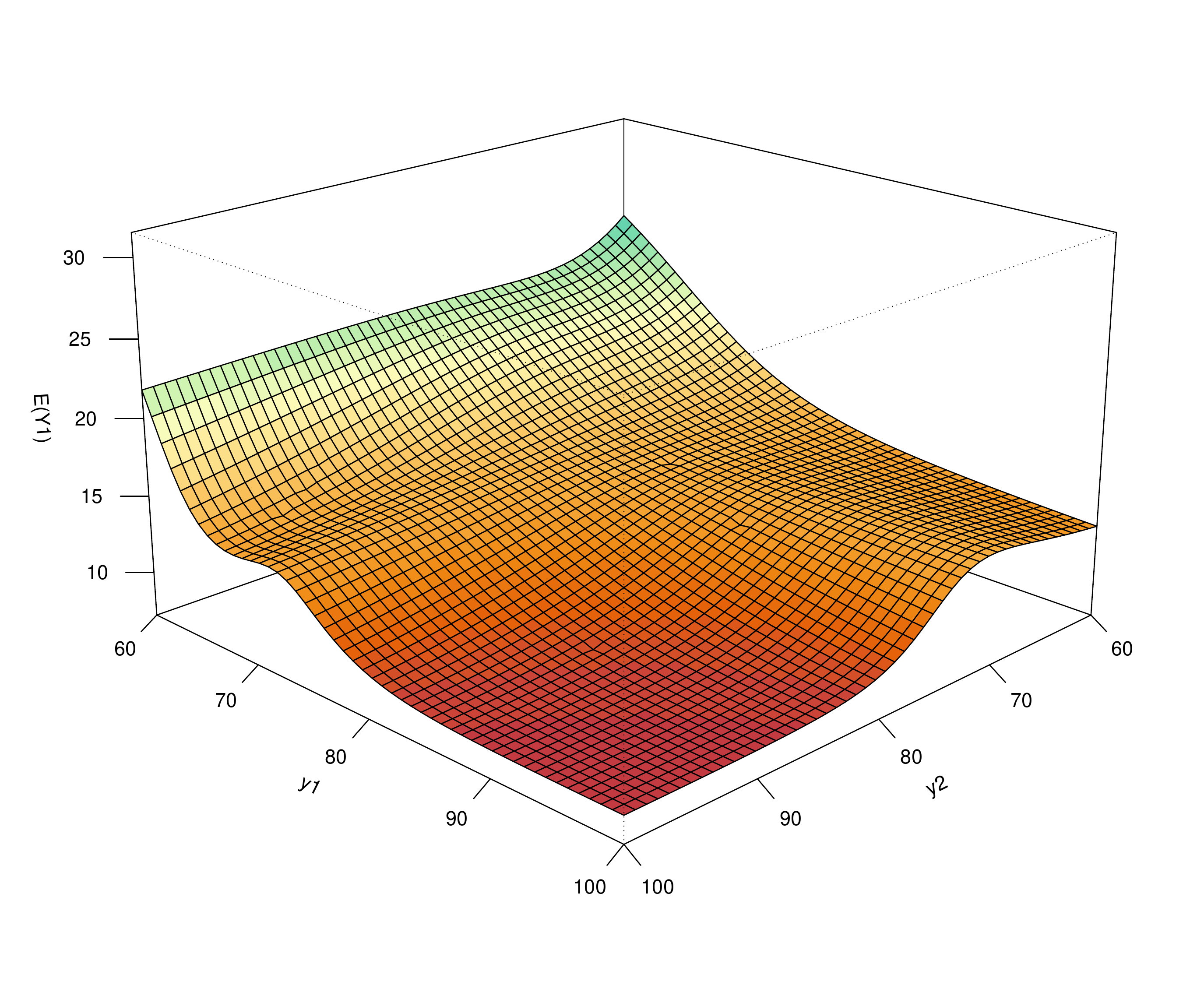}
        \caption{Man.}
    \end{subfigure}
    \hfill
        \begin{subfigure}[b]{0.495\textwidth}
        \includegraphics[width=1\textwidth]{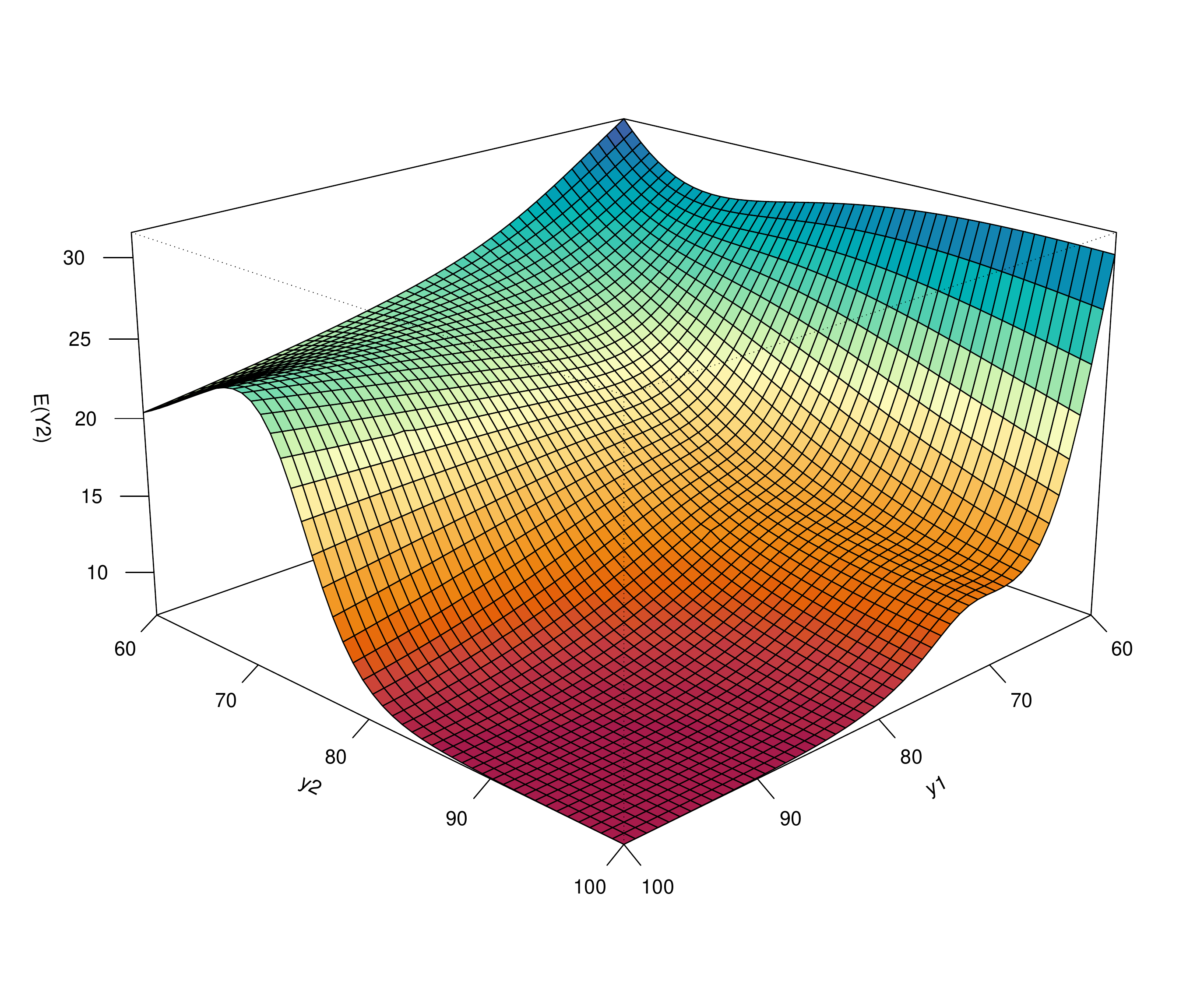}
        \caption{Woman.}
    \end{subfigure}
    \caption{Marginal expected excess survival times as a function of men's age $y_1$ and women's $y_2$ at issue.}
    \label{fig: exp 3d}
\end{figure}

Let us finally also use the model fit to get some insight into expected remaining lifetimes in the couple. Using Property \eqref{cond_dist_surv}, we are able to derive expected survival times for an individual, conditional on the survival time of their partner. Moreover, using optimal regression coefficients found by Algorithm \ref{alg1} (Table \ref{table: coef}) we can study how marginal expectations vary with ages in a couple. Letting the man's and woman's age vary from age $60$ to $100$, we get for each age combination a distinct mIPH distribution for the random vector $\mat{Y}$. Given these distributions, we summarise in Figure \ref{fig: exp 3d} how the expected remaining lifetime at issue changes as a function of the male's and female's age at that point in time. For both man and woman, the marginal expected survival times decrease when both individuals in the couple grow older. For any specific age, both men and women have larger marginal expectations as their spouses become younger. That marginal expectation varies more for women, as one can see from the steepness and range of the women's curve in Figure \ref{fig: exp 3d}. Another way to interpret this is that men's expected survival times are affected less by the age of their partner, compared to women.\\

\begin{figure}[ht]
    \centering
    \begin{subfigure}[b]{0.495\textwidth}
        \includegraphics[width=1\textwidth]{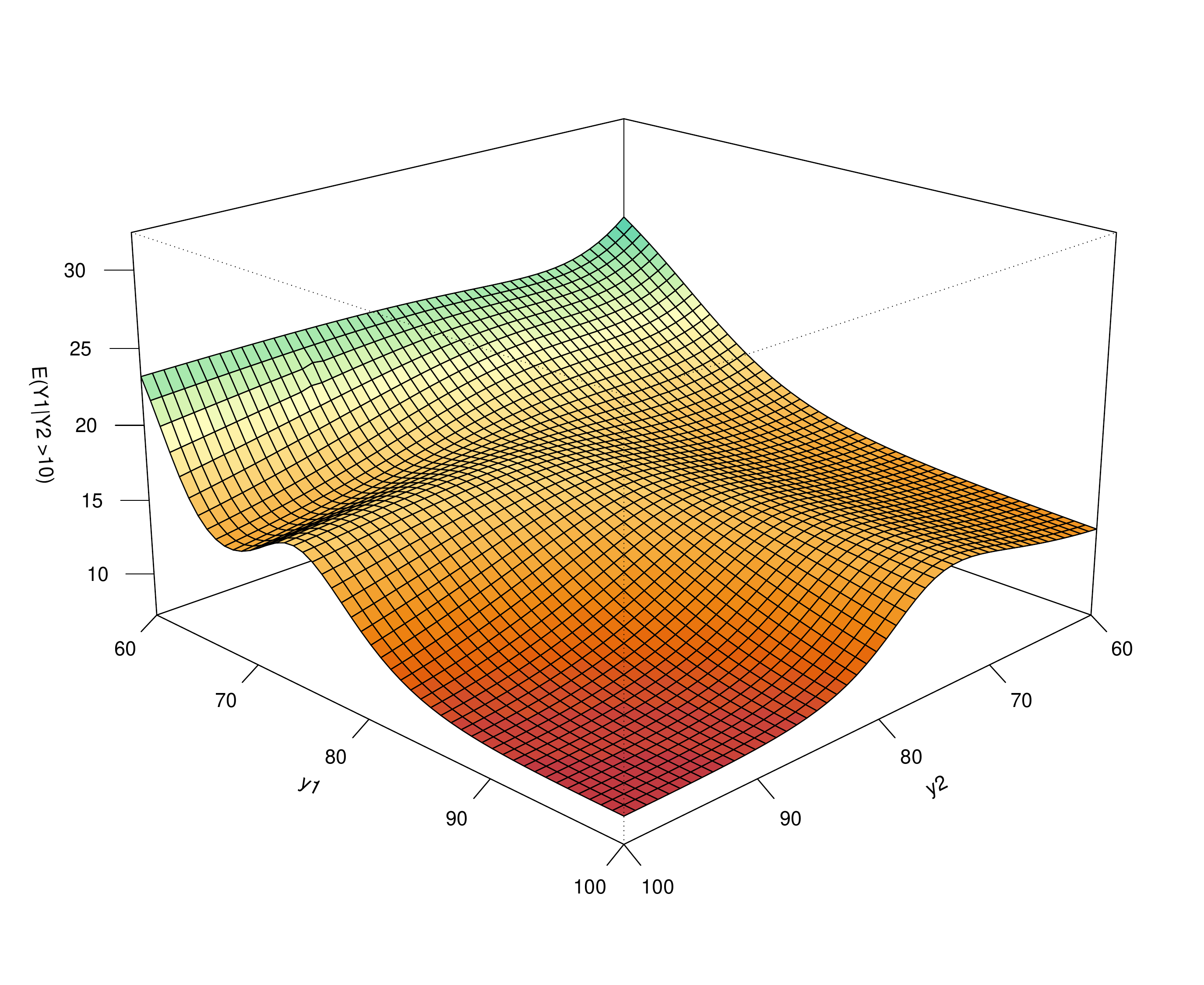}
        \caption{Man.}
    \end{subfigure}
    \hfill
        \begin{subfigure}[b]{0.495\textwidth}
        \includegraphics[width=1\textwidth]{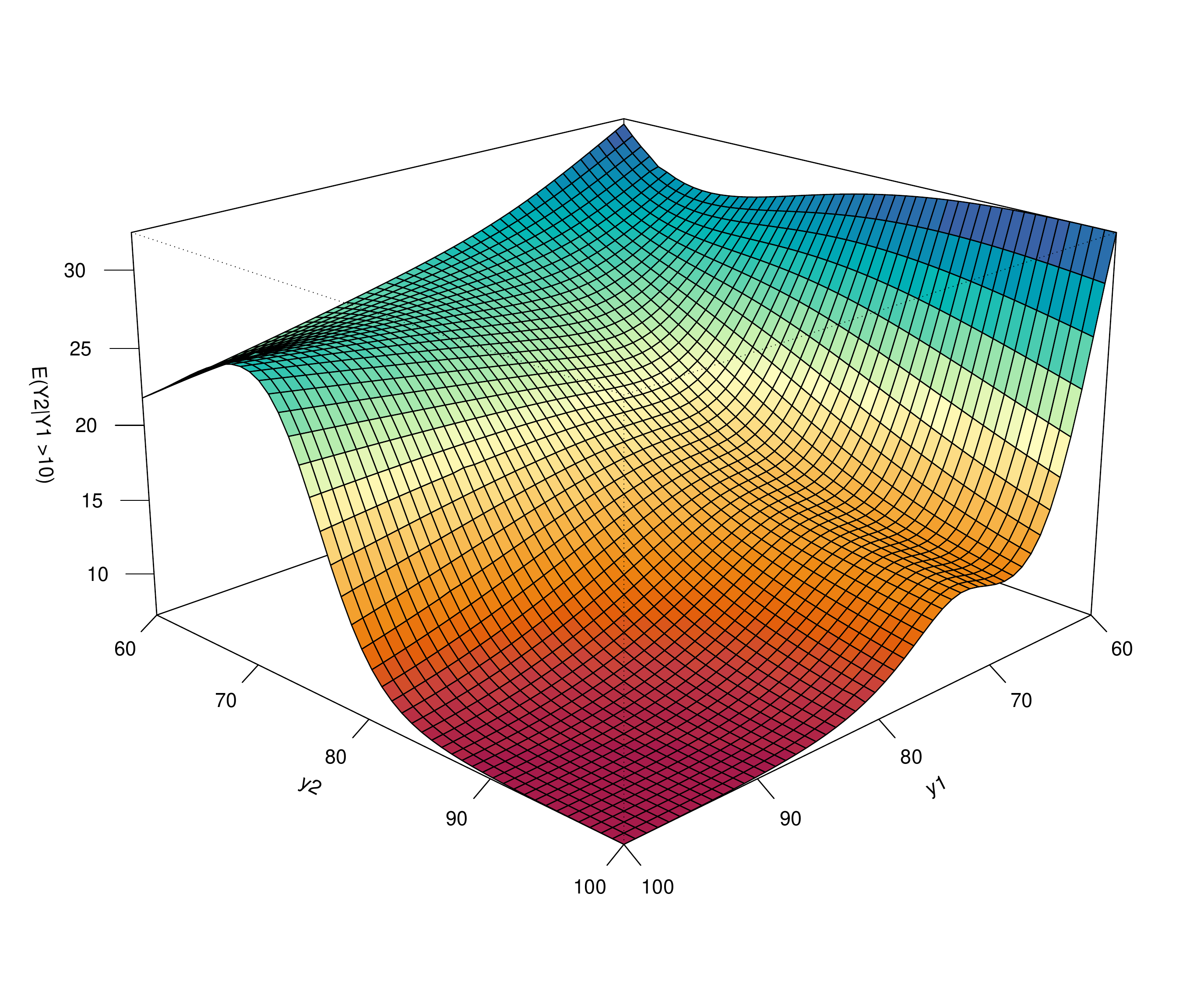}
        \caption{Woman.}
    \end{subfigure}
    \caption{Marginal expected excess survival times, conditional on spouse survival during 10 years.}
    \label{fig: exp_10 3d}
\end{figure}
\begin{figure}[ht]
    \centering
    \begin{subfigure}[b]{0.495\textwidth}
        \includegraphics[width=1\textwidth]{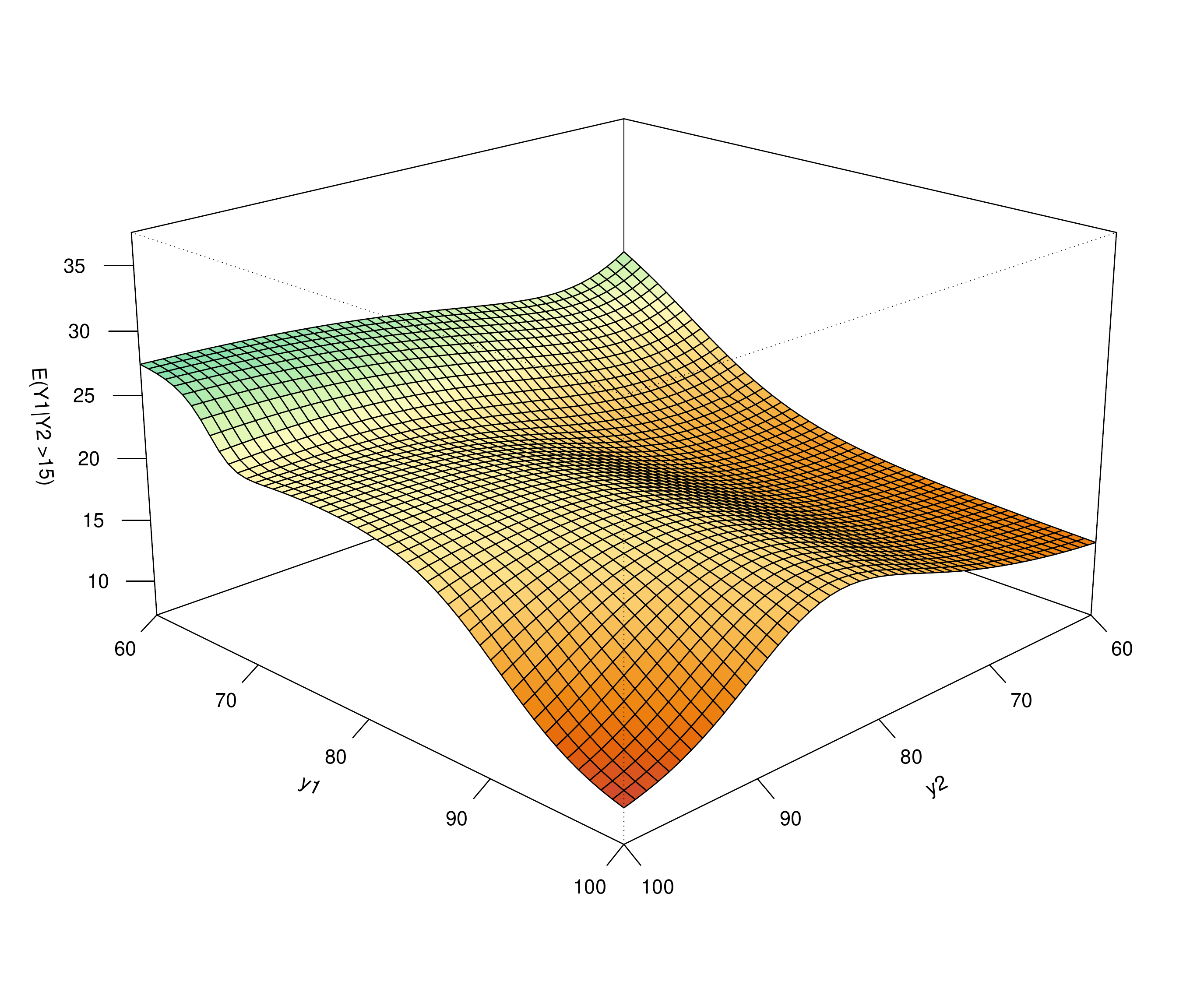}
        \caption{Man.}
    \end{subfigure}
    \hfill
        \begin{subfigure}[b]{0.495\textwidth}
        \includegraphics[width=1\textwidth]{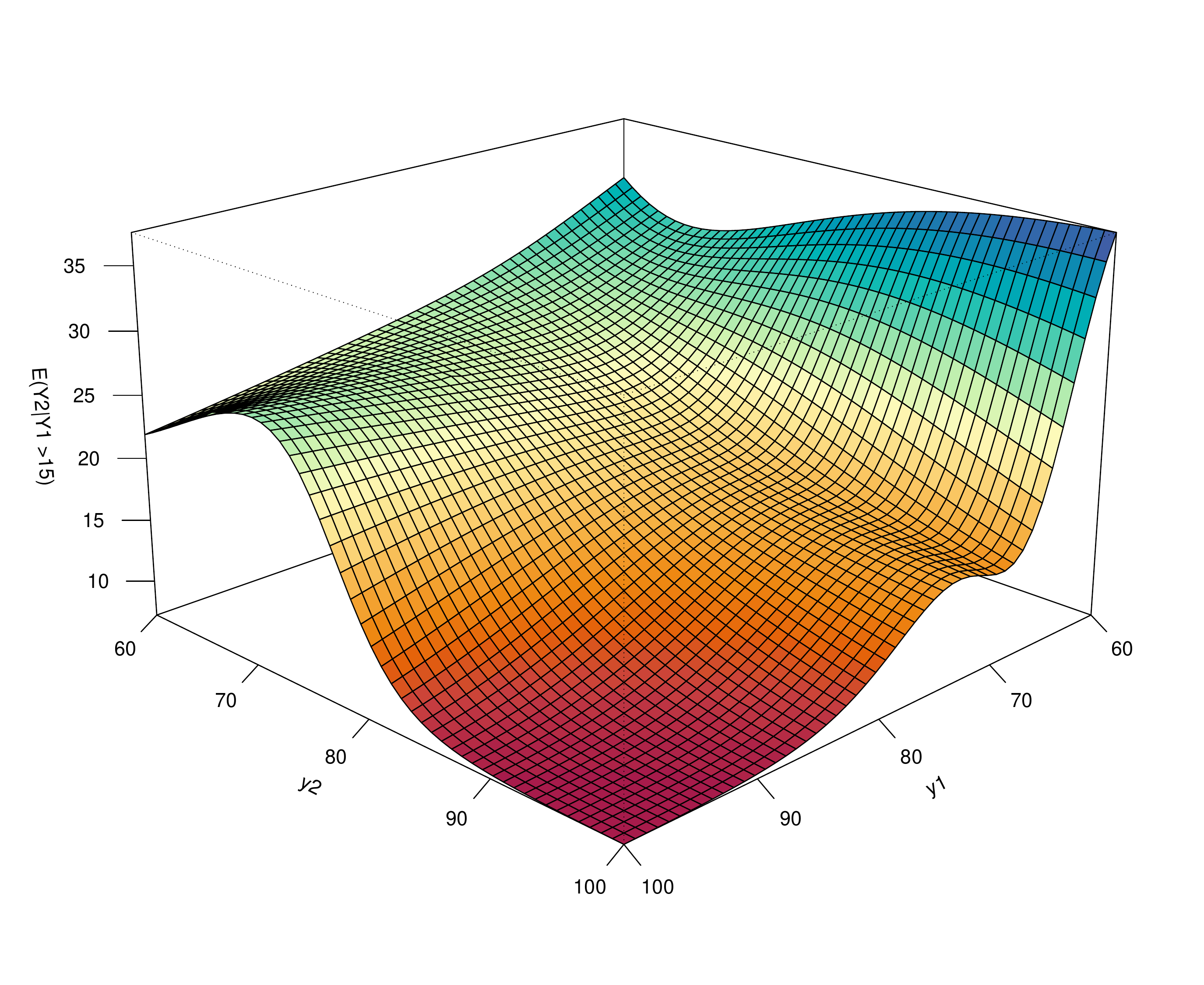}
        \caption{Woman.}
    \end{subfigure}
    \caption{Marginal expected excess survival times, conditional on spouse survival during 15 years.}
    \label{fig: exp_15 3d}
\end{figure}
Besides age, conditioning also on the survival time of the spouse affects the initial distribution vector, leading to different expected values. Figures \ref{fig: exp_10 3d} and \ref{fig: exp_15 3d} present marginal expected survival times, conditional on a spouse survival of at least $10$ and $15$ years counted from issue of the policy, respectively. 
The respective change in the men's remaining life expectation is notable by mere visual comparison, whereas for women it is less pronounced. From Figure \ref{fig: exp_10 3d} we see that the expectation increases for almost all ages, compared with Figure \ref{fig: exp 3d}. For men, one notices that the shape of the curve is changed for ages $65\le y_1\le75$. Moreover, for older ages $y_1$ and $y_2$ the curve is now slightly steeper than before. In the women's case, marginal expectations for a woman aged $60$ or $100$ with husband aged $60$ are now equivalent, and no other major change can be easily observed. The marginal expectation for men whose spouses survive at least 15 years are very much different from their unconditional counterparts. Inspecting Figure \ref{fig: exp_15 3d}, one can see a sizeable change of the curve. Men with spouses of age $y_2\in(80,100)$ are now expected to survive much longer than before, while once again for women we only have a rather minor twist of the curve.
\textcolor{black}{
In all figures we see that the women's curves do not show major change. Still, they suggest that women are more sensitive to both age and survival of their partner. 
}
%This may suggest that men's survival is more sensitive to their spouses' survival times than vice versa.  
%%%%%%%%%%%%%%%%%%%%%%%%%%%%%%%%%%%%%%%%%%%%%%%%%%%%%%%%%%%%%%%%%%
%----------------------------------------------------------------%
%%%%%%%%%%%%%%%%%%%%%%%%%%%%%%%%%%%%%%%%%%%%%%%%%%%%%%%%%%%%%%%%%%
\section{Conclusion}
In this paper, we introduce the mIPH class, study some of its properties and develop an estimation procedure that allows for right-censored observations and covariate information. In particular, we use this framework to propose a bivariate Matrix-Gompertz distribution for the modelling of excess joint lifetimes of couples. 
Adapting a respective Expectation-Maximisation algorithm, we estimate sub-intensity matrices, inhomogeneity functions and different initial distribution vectors without separating joint features from marginals. Initial probabilities are assumed to be linked to spouses' ages at the issue of an insurance policy. Employing multinomial logistic regressions to predict the latter, tailor-made bivariate distributions are produced that reflect distinct ageing dynamics and dependence structures. 
The resulting mIPH distributions showcase strong positive concordance for remaining lifetimes of spouses, particularly when the difference in age at the issue of the policy is small. 

The results and illustrations given in this paper demonstrate the accuracy and flexibility of the mIPH class, which may also be employed in areas beyond the present lifetime setup, including applications in non-life insurance. \textcolor{black}{The mIPH class may be considered a valid alternative to copula-based methods}, particularly when one wants to estimate marginal and multivariate properties at the same time, and has sufficiently many data points available to keep the pre-imposed dependence assumptions (and structure) minimal. In addition, modelling with members of the mIPH class allows for an immediate causal interpretation of the resulting model in terms of common ageing through stages. 

\textbf{Acknowledgement.} We would like to thank the Society of Actuaries, through the courtesy of Edward W. Frees and Emiliano Valdez for access to the data set used in this paper. Financial support from the Swiss National Science Foundation Project 200021\_191984 is gratefully acknowledged.

\bibliography{jointmiph.bib}

\begin{thebibliography}{10}

\bibitem{albrecher2019inhomogeneous}
H.~Albrecher and M.~Bladt.
\newblock Inhomogeneous phase-type distributions and heavy tails.
\newblock {\em Journal of Applied Probability}, 56(4):1044--1064, 2019.

\bibitem{ABBY21}
H.~Albrecher, M.~Bladt, M.~Bladt, and J.~Yslas.
\newblock Mortality modeling and regression with matrix distributions.
\newblock {\em Insurance: Mathematics and Economics}, 107:68--87, 2022.

\bibitem{albrecher2022penalised}
H.~Albrecher, M.~Bladt, and A.~J.~A. M\"uller.
\newblock Penalised likelihood methods for phase-type dimension selection.
\newblock {\em Preprint, University of Lausanne}, 2022+.

\bibitem{ABY22}
H.~Albrecher, M.~Bladt, and J.~Yslas.
\newblock Fitting inhomogeneous phase-type distributions to data: the
  univariate and the multivariate case.
\newblock {\em Scandinavian Journal of Statistics}, 49(1):44--77, 2022.

\bibitem{Asmussen2019}
S.~Asmussen, P.~J. Laub, and H.~Yang.
\newblock Phase-type models in life insurance: Fitting and valuation of
  equity-linked benefits.
\newblock {\em Risks}, 7(1), 2019.

\bibitem{asmussen(1996)em}
S.~Asmussen, O.~Nerman, and M.~Olsson.
\newblock Fitting phase-type distributions via the em algorithm.
\newblock {\em Scandinavian Journal of Statistics}, 23(4):419--441, 1996.

\bibitem{bladt2022tractable}
M.~Bladt.
\newblock A tractable class of multivariate phase-type distributions for loss
  modeling.
\newblock {\em Preprint, University of Lausanne, arXiv preprint
  arXiv:2110.05179}, 2021.

\bibitem{Bladt2017}
M.~Bladt and B.~F. Nielsen.
\newblock {\em Matrix-Exponential Distributions in Applied Probability},
  volume~81.
\newblock Springer, 2017.

\bibitem{BladtYslas2022}
M.~Bladt and J.~Yslas.
\newblock Phase-type mixture-of-experts regression for loss severities.
\newblock {\em Scandinavian Actuarial Journal}, 0(0):1--27, 2022.

\bibitem{carriere2000bivariate}
J.~F. Carriere.
\newblock Bivariate survival models for coupled lives.
\newblock {\em Scandinavian Actuarial Journal}, 2000(1):17--32, 2000.

\bibitem{clayton1978model}
D.~G. Clayton.
\newblock A model for association in bivariate life tables and its application
  in epidemiological studies of familial tendency in chronic disease incidence.
\newblock {\em Biometrika}, 65(1):141--151, 1978.

\bibitem{dabrowska1989}
D.~M. Dabrowska.
\newblock Uniform consistency of the kernel conditional kaplan-meier estimate.
\newblock {\em The Annals of Statistics}, 17(3):1157--1167, 1989.

\bibitem{dufresne(2018)agediff}
F.~Dufresne, E.~Hashorva, G.~Ratovomirija, and Y.~Toukourou.
\newblock On age difference in joint lifetime modelling with life insurance
  annuity applications.
\newblock {\em Annals of Actuarial Science}, 12(2):350--371, 2018.

\bibitem{faddy2002penalised}
M.J. Faddy.
\newblock Penalised maximum likelihood estimation of the parameters in a
  {C}oxian phase-type distribution.
\newblock In {\em Matrix-analytic methods: theory and applications}, pages
  107--114. World Scientific, 2002.

\bibitem{frees(1996)}
E.~W. Frees, J.~Carriere, and E.~Valdez.
\newblock Annuity valuation with dependent mortality.
\newblock {\em The Journal of Risk and Insurance}, 63(2):229--261, 1996.

\bibitem{gobbi_kolev_mulinacci_2019}
F.~Gobbi, N.~Kolev, and S.~Mulinacci.
\newblock Joint life insurance pricing using extended marshall-olkin models.
\newblock {\em ASTIN Bulletin}, 49(2):409--432, 2019.

\bibitem{markovapproach2011}
M.~Ji, M.~Hardy, and J.~S.-H. Li.
\newblock Markovian approaches to joint-life mortality.
\newblock {\em North American Actuarial Journal}, 15(3):357--376, 2011.

\bibitem{markovageing2007Lin&Liu}
X.~Lin, X. S.and~Liu.
\newblock Markov aging process and phase-type law of mortality.
\newblock {\em North American Actuarial Journal}, 11(4):92--109, 2007.

\bibitem{luciano2008}
J.~Luciano, E.and~Spreeuw and E.~Vigna.
\newblock Modelling stochastic mortality for dependent lives.
\newblock {\em Insurance: Mathematics and Economics}, 43(2):234--244, 2008.

\bibitem{sarmanov2021}
K.~Moutanabbir and H.~Abdelrahman.
\newblock Bivariate {S}armanov phase-type distributions for joint lifetimes
  modeling.
\newblock {\em Methodology and Computing in Applied Probability},
  24(2):1093--1118, 2022.

\bibitem{olsson(1996)rcens}
M.~Olsson.
\newblock Estimation of phase-type distributions from censored data.
\newblock {\em Scandinavian Journal of Statistics}, 23(4):443--460, 1996.

\bibitem{shemyakin2006}
A.~E. Shemyakin and H.~Youn.
\newblock Copula models of joint last survivor analysis.
\newblock {\em Applied Stochastic Models in Business and Industry},
  22(1):211--224, 2006.

\bibitem{spreeuw_owadally_2013}
J.~Spreeuw and I.~Owadally.
\newblock Investigating the broken-heart effect: a model for short-term
  dependence between the remaining lifetimes of joint lives.
\newblock {\em Annals of Actuarial Science}, 7(2):236–257, 2013.

\end{thebibliography}

\end{document}